\documentclass[conf]{new-aiaa}
\usepackage[utf8]{inputenc}
\usepackage{subfigure}
\newtheorem{theorem}{Theorem}
\usepackage{graphicx}
\usepackage{amsmath}
\usepackage[version=4]{mhchem}
\usepackage{siunitx}
\usepackage[utf8]{inputenc}
 \usepackage[usenames, dvipsnames]{color}
\usepackage{nomencl}
\makenomenclature
 \usepackage{longtable,tabularx}
\title{Multi-UAV Continuum
Deformation Flight Optimization in Cluttered Urban Environments}

\author{Hossein Rastgoftar\footnote{Assistant Research Scientist, Department of Aerospace Engineering, University  of  Michigan   (e-mail:  hosseinr@umich.edu).} and Ella M. Atkins\footnote{Professor, Department of Aerospace Engineering, Univer-sity  of  Michigan (e-mail:  ematkins@umich.edu).}}
\affil{Department of Aerospace Engineering, University  of  Michigan,  Ann  Arbor,  MI,  48109  USA }

\begin{document}

\maketitle

\begin{abstract}
This paper studies collective motion optimization of a fleet of UAVs flying over a populated and geometrically constrained area. The paper treats UAVs as particles of a deformable body, thus, UAV coordination is defined by a homeomorphic continuum deformation function.  
Under continuum deformation, the distance between individual UAVs can significantly change while assuring the UAVs don’t collide, enabling a swarm to travel through the potentially cluttered environment. To ensure inter-agent and obstacle collision avoidance, the paper formulates safety requirements as inequality constraints of the coordination optimization problem.  The main objective of the paper is then to optimize continuum deformation of the UAV team satisfying all continuum deformation inequality constraints. Given initial and target configurations, the cost is defined as a weighted sum of the travel distance and distributed cost proportional to the likelihood of human presence. 
\end{abstract}

\printnomenclature


\section{Introduction}
Multi-agent coordination has been widely studied over the past two decades. Formation flight offers several advantages such as failure resilience \cite{rieger2013resilient} and reduced mission cost \cite{zhao2013energy}. Multi-agent coordination applications include but are not limited to surveillance \cite{botts2016multi}, air traffic  management \cite{zhu2015junction}, formation flight \cite{oh2015survey}, and connected vehicle control \cite{feng2015real}.  


{\color{black}
Extensive previous work has been undertaken for multi-agent coordination with application to ground and air vehicles.  Virtual structure, consensus, containment control, and continuum deformation offer agent coordination in a $3D$ motion space. Virtual structure formulations coordinate agents in a centralized fashion. In a virtual structure, each agent's desired position consists of a reference position vector and a relative displacement vector with respect to the reference \cite{low2011flexible}. If the agents' relative distances from a reference position remain constant, the multi-agent system can be treated as a rigid body \cite{li2008formation}. A flexible virtual structure formulation is also studied in \cite{essghaier2011co}. 
Consensus \cite{ren2009distributed, ren2007information, ding2013network}, containment \cite{li2015containment, ji2008containment}, and continuum deformation \cite{rastgoftar2016continuum, rastgoftar2016asymptotic, rastgoftar2014evolution, rastgoftar2017continuum, rastgoftar2015swarm} are decentralized multi-agent coordination approaches. Consensus is perhaps the most common approach for formation and cooperative control. Both leaderless \cite{ren2009distributed}  and leader-based  \cite{ding2013network} consensus approaches have been proposed. Consensus under switching communication topologies is studied in Refs. \cite{cheng2014mean, ma2015centralized, ni2010leader}, while stability of consensus coordination is analyzed in Refs. \cite{hou2017consensus, nazari2016decentralized}. 

With containment control, leaders move independently and guide motions of the agent team. Followers communicate with select in-neighbor agents to acquire coordination through local communication \cite{li2015containment, ji2008containment}. Containment control is studied in Refs. \cite{cao2012distributed} and \cite{cao2009containment}.  Retarded containment control stability is analyzed in Ref. \cite{li2013distributed}. Finite-time \cite{meng2010distributed} and heterogeneous agent \cite{zheng2014containment} containment control formulations have also been developed.

Under continuum deformation, inter-agent distances can change significantly while inter-agent collision is avoided \cite{lal2006continuum, lal2006new}.  Coordination is  formulated as a decentralized leader-follower formation control problem in Ref. \cite{rastgoftar2014evolution}. Ref. \cite{rastgoftar2014evolution} formulates an $n$-dimensional ($n=1,2,3$) homogeneous transformation based on the trajectories of $n+1$ leaders forming an $n$-dimensional polytope in a $3D$ motion space, while Ref. \cite{rastgoftar2014evolution} shows how follower agents can acquire desired trajectories through local communication. Decentralized continuum deformation coordination using an area preservation and alignment strategy is demonstrated in \cite{rastgoftar2016asymptotic} and \cite{rastgoftar2015swarm}, while Ref. \cite{rastgoftar2016continuum} investigates the stability of continuum deformation coordination with communication delay. Sufficient conditions for safe continuum deformation coordination and inter-agent collision avoidance are developed in Refs. \cite{rastgoftar2016continuum, rastgoftar2016asymptotic}. Ref. \cite{rastgoftar2017continuum} formulates continuum deformation coordination under switching communication topologies.

Robot motion planning has been widely studied in the literature. A* \cite{stentz1994optimal} and dynamic programming \cite{roozegar2016optimal} support globally-optimal planning over a discrete grid or predefined waypoint set.  Rapidly-expanding Random Trees
(RRT) \cite{melchior2007particle} and reachability graphs \cite{liu1990flexible}  are available 
graph-based methods for real-time motion planning.  In Ref. 
\cite{kuffner2000rrt}  Gaussian process (GP) and RRT are applied 
build real-time motion plans. Model predictive control (MPC) \cite{camacho2013model} is a well-known approach for real-time optimization as well as real-time path planning in an obstacle laden environment \cite{wang2007cooperative}. 
}

This paper advances the authors' previous continuum deformation contributions for single/double integrator agents to safe continuum deformation optimization of UAVs with nonlinear dynamics. We consider coordination of a UAV team forming a triangle in a $2D$ motion plane, called a \textit{leading triangle}. UAV coordination is guided by three leaders at the vertices of the leading triangle and acquired by followers through direct communication with leaders. Compared to previous work, we offer a novel contribution on continuum deformation optimization of a UAV team over a populated and geometrically-constrained environment by incorporating obstacle (no-fly-zone) and population risk metrics to the formulation. While inter-agent and obstacle collision avoidance are guaranteed, UAV coordination is optimized by minimizing travel distance and time of flight over populated areas.  

This paper is organized as follows. Theoretical background in Section \ref{Preliminaries} is followed by a review of multi-quadcopter system (MQS) continuum deformation in Section \ref{Continuum Deformation Coordination}.  MQS continuum deformation optimization is mathematically defined in Section \ref{Problem Statement}. The path planning optimization strategy in Section \ref{PathPlanning} is followed by MQS evolution analysis in Section \ref{Continuum Deformation Acquisition}. UAV dynamics and control are modeled in Sections  \ref{UAV Dynamics} and  \ref{UAV Control}, respectively. Simulation results in Section \ref{Simulation Results}  are followed by concluding remarks in Section \ref{Conclusion}.

\section{Preliminaries}
\label{Preliminaries}

\subsection{Coordinate Systems} We consider an inertial (or ground) coordinate system with the bases $\hat{\mathbf{e}}_1$, $\hat{\mathbf{e}}_2$, $\hat{\mathbf{e}}_3$. Note that  $\hat{\mathbf{e}}_1$, $\hat{\mathbf{e}}_2$, $\hat{\mathbf{e}}_3$ are fixed on the ground. Furthermore, each UAV has it{\color{black}s} own local coordinate system, called \textit{body coordinates}. Mutually perpendicular unit vectors $\mathbf{i}_{b,i}$, $\mathbf{j}_{b,i}$, and $\mathbf{k}_{b,i}$ are the bases of UAV $i$ body coordinate system.  $\mathbf{i}_{b,i}$, $\mathbf{j}_{b,i}$, and $\mathbf{k}_{b,i}$ are related to  $\hat{\mathbf{e}}_1$, $\hat{\mathbf{e}}_2$, $\hat{\mathbf{e}}_3$ by
\begin{subequations}
\begin{equation}
    \begin{bmatrix}
    \hat{\mathbf{i}}_{b,i}\\
    \hat{\mathbf{j}}_{b,i}\\
    \hat{\mathbf{k}}_{b,i}\\
    \end{bmatrix}
    =
    \mathcal{R}_{{\phi_i}\theta_i\psi_i}
    \begin{bmatrix}
    \hat{\mathbf{e}}_1\\
    \hat{\mathbf{e}}_2\\
    \hat{\mathbf{e}}_3\\
    \end{bmatrix}
\end{equation}
\begin{equation}
\begin{split}
     \mathcal{R}_{{\phi_i}\theta_i\psi_i}=
    \begin{bmatrix}
    C_{\theta_i} C_{\psi_i}&C_{\theta_i} S_{\psi_i}&-S_{\theta_i}\\
    S_{\phi_i} S_{\theta_i} C_{\psi_i}-C_{\phi_i} S_{\psi_i}& S_{\phi_i} S_{\theta_i} S_{\psi_i}+C_{\phi_i} C_{\psi_i}&S_{\phi_i} C_{\theta_i}\\
    C_{\phi_i} S_{\theta_i} C_{\psi_i}+S_{\phi_i} S_{\psi_i}& C_{\phi_i} S_{\theta_i} S_{\psi_i}-S_{\phi_i} C_{\psi_i}&C_{\phi_i} C_{\theta_i}
    \end{bmatrix}
    ,
\end{split}
\end{equation}
\end{subequations}
where $C_{\left(\cdot\right)}$ and $S_{\left(\cdot\right)}$ abbreviate $\cos\left(\cdot\right)$ and $\sin\left(\cdot\right)$, respectively. Furthermore,   $\phi_i$, $\theta_i$, and $\psi_i$ are the roll, pitch, and yaw angles of UAVs $i\in \mathcal{V}$.
\nomenclature{$\hat{\mathbf{e}}_1,~\hat{\mathbf{e}}_2,~\hat{\mathbf{e}}_3$}{Bases of the ground coordinate system}
\nomenclature{$\hat{\mathbf{i}}_{b,i},~\hat{\mathbf{j}}_{b,i},~\hat{\mathbf{j}}_{b,i}$}{Bases of UAV $i$ body frame}
\nomenclature{$\phi_i,~\theta_i,~\psi_i$}{Euler angles of UAV $i$}
\nomenclature{$\mathbf{r}_i$}{Actual position of UAV $i$}
\nomenclature{$x_i,~y_i,~z_i$}{Actual position components}
\nomenclature{$\mathcal{V}_F$}{Follower set}
\nomenclature{$\mathcal{V}_L$}{Leader set}
\nomenclature{$\mathcal{V}$}{Set of UAV index numbers}
\subsection{Position Terminologies} 
Throughout the paper all position vectors are expressed with respect to the ground coordinate system with the bases $\hat{\mathbf{e}}_1$, $\hat{\mathbf{e}}_2$, $\hat{\mathbf{e}}_3$.  
\begin{equation}
   i\in\mathcal{V},\qquad \mathbf{r}_i=x_i\hat{\mathbf{e}}_1+y_i\hat{\mathbf{e}}_2+z_i\hat{\mathbf{e}}_3.
\end{equation}
\nomenclature{$\mathbf{R}_{i,0}$}{Initial position of UAV $i$}
\nomenclature{$\mathbf{R}_{i,F}$}{Target position of UAV $i$}
\nomenclature{$\mathbf{r}_{i,HT}$}{Global desired position of UAV $i$}
denotes the \textbf{actual position} of UAV $i\in \mathcal{V}$.
\begin{equation}
\label{riht}
i\in \mathcal{V},\qquad \mathbf{r}_{i,HT}=x_{i,HT}\hat{\mathbf{e}}_1+y_{i,HT}\hat{\mathbf{e}}_2+z_{i,HT}\hat{\mathbf{e}}_3
\end{equation}
denotes the \textbf{global desired position} of agent $i\in \mathcal{V}$. Note that $\mathbf{r}_{i,HT}$, defined by a class of continuum deformation mappings called \textit{homogeneous transformation}, is formulated based on leaders' trajectories in Section \ref{Continuum Deformation Coordination}. 
\begin{equation}
    i\in\mathcal{V},\qquad \mathbf{R}_{i,0}=X_{i,0}\hat{\mathbf{e}}_1+Y_{i,0}\hat{\mathbf{e}}_2+Z_{i,0}\hat{\mathbf{e}}_3
\end{equation}
and
\begin{equation}
    i\in\mathcal{V},\qquad\mathbf{R}_{i,F}=X_{i,F}\hat{\mathbf{e}}_1+Y_{i,F}\hat{\mathbf{e}}_2+Z_{i,F}\hat{\mathbf{e}}_3
\end{equation}
denote \textbf{initial position} and \textbf{target position} of UAV $i\in \mathcal{V}$. 

\section{Continuum Deformation Coordination Definition}
\label{Continuum Deformation Coordination}
 Consider a group of $N$ UAVs moving in a $2D$ motion space. UAVs are identified by index numbers $1$ through $N$ defined by the set $\mathcal{V}$, e.g. $\mathcal{V}=\{1,\cdots,N\}$. UAVs are enclosed by a triangular domain, called \textit{leading triangle}. MQS collective dynamics is guided by three leaders with identification numbers $1$, $2$, and $3$ defined by set $\mathcal{V}_L$, e.g. $\mathcal{V}_L=\{1,2,3\}$. The remaining UAVs inside the leading triangle are followers. Follower UAVs' index numbers are defined by the set $\mathcal{V}_F=\mathcal{V}\setminus \mathcal{V}_L$. Global desired position of UAV $i\in \mathcal{V}$ is defined by a homogeneous transformation 
\begin{equation}
\label{maincontinum}
    i\in \mathcal{V},t\geq t_0\qquad \mathbf{r}_{i,HT}\left(t\right)=Q\left(t\right)\mathbf{R}_{i,0}+\mathbf{D}\left(t\right)
\end{equation}
where $Q\in \mathbb{R}^{3\times 3}$ is the continuum deformation Jacobian matrix and $\mathbf{D}(t)$ is a rigid-body displacement vector. Because this paper studies $2D$ continuum deformation in the $X-Y$ plane,
\begin{subequations}
\begin{equation}
    Q=
    \begin{bmatrix}
    Q_{\mathrm{CD}}&\mathbf{0}\\
    \mathbf{0}&1
    \end{bmatrix}
    =
    \begin{bmatrix}
    Q_{1,1}&Q_{1,2}&0\\
    Q_{2,1}&Q_{2,2}&0\\
    0&0&1
    \end{bmatrix}
\end{equation}
\begin{equation}
    \mathbf{D}=
    \begin{bmatrix}
    D_1&D_2&0
    \end{bmatrix}
    ^T.
\end{equation}
\end{subequations}
Because the $z$ component of $\mathbf{D}$ is $0$, ${z}_{i,HT}\left(t\right)={Z}_{i,0}$ ($\forall i\in \mathcal{V},\forall t$). This paper assumes that the $z$ components of the agents' global desired positions are the same:
\begin{equation}
    \forall i\in\mathcal{V},\forall t,\qquad z_{i,HT}=z_{HT}.
\end{equation}

Leaders form a triangle at all times $t\geq t_0$, therefore, 
\begin{equation}
\label{leaderrankk}
    \forall t\geq t_0,\qquad \mathrm{rank}\left(
    \begin{bmatrix}
    \mathbf{r}_{2,HT}-\mathbf{r}_{1,HT}&\mathbf{r}_{3,HT}-\mathbf{r}_{1,HT}
    \end{bmatrix}
    \right)
    =2.
\end{equation}
Because the rank condition \eqref{leaderrankk} is satisfied at initial time $t_0$, elements of $Q_{\mathrm{CD}}$,  $D_1$ and $D_2$ are uniquely related to leaders' global desired positions components:
\begin{equation}
\begin{bmatrix}
Q_{1,1}(t)\\
Q_{1,2}(t)\\
Q_{2,1}(t)\\
Q_{2,2}(t)\\
D_1(t)\\
D_2(t)
\end{bmatrix}
=
\begin{bmatrix}
X_{1,0}&Y_{1,0}&0&0&1&0\\
X_{2,0}&Y_{2,0}&0&0&1&0\\
X_{3,0}&Y_{3,0}&0&0&1&0\\
0&0&X_{1,0}&Y_{1,0}&0&1\\
0&0&X_{2,0}&Y_{2,0}&0&1\\
0&0&X_{3,0}&Y_{3,0}&0&1\\
\end{bmatrix}
\begin{bmatrix}
x_{1,HT}(t)\\
x_{2,HT}(t)\\
x_{3,HT}(t)\\
y_{1,HT}(t)\\
y_{2,HT}(t)\\
y_{3,HT}(t)\\
\end{bmatrix}
.
\end{equation}

\nomenclature{$Q$}{Homogeneous deformation Jacobian matrix}
\nomenclature{$\mathbf{D}$}{Rigid body displacement vector}
 Using polar decomposition, $Q_{\mathrm{CD}}$ can be expressed as
\begin{equation}
Q_{\mathrm{CD}}=\mathcal{R}_{\mathrm{CD}}U_{\mathrm{CD}}
\end{equation}
\nomenclature{$\mathcal{R}_D$}{Homogeneous deformation rotation matrix}
\nomenclature{$\mathcal{U}_D$}{Homogeneous deformation pure deformation matrix matrix}
where $U_{\mathrm{CD}}$ is a positive definite (and symmetric) matrix and $R_{\mathrm{CD}}$ is an orthogonal matrix, e.g $\mathcal{R}_{\mathrm{CD}}^T\mathcal{R}_{\mathrm{CD}}=I_2$. Eigenvalues of the matrix $U_{\mathrm{CD}}$ are positive and real and denoted by $\lambda_1$ and $\lambda_2$ ($0<\lambda_1\leq \lambda_2$). 

\textbf{Key Property of a Homogeneous Deformation:} Let leaders form a triangle at all times $t$. Therefore,
\[
    \forall t\geq t_0,~ \mathrm{Rank}\left(
    \begin{bmatrix}
    \mathbf{r}_{2,HT}-\mathbf{r}_{1,HT}&\mathbf{r}_{3,HT}-\mathbf{r}_{1,HT}
    \end{bmatrix}
    \right)
    =2
    .
\]

Under a homogeneous deformation, $X$ and $Y$ components of the global desired position of UAV $i\in \mathcal{V}$ can be expressed as \cite{rastgoftar2016continuum}
\begin{equation}
\label{globdespos}
    \begin{bmatrix}
     x_{i,HT}(t)\\
     y_{i,HT}(t)\\
    \end{bmatrix}
    =\sum_{j=1}^3
    \alpha_{i,j}
    \begin{bmatrix}
     x_{j,HT}(t)\\
     y_{j,HT}(t)\\
    \end{bmatrix}
    ,
\end{equation}
where $\alpha_{i,1}$, $\alpha_{i,2}$, are $\alpha_{i,3}$ \textbf{time-invariant} parameters and
\begin{equation}
    \alpha_{i,1}+\alpha_{i,2}+\alpha_{i,3}=1.
\end{equation}
Parameters $\alpha_{i,1}$, $\alpha_{i,2}$, and $\alpha_{i,3}$ are computed from the initial position of UAV $i$ and the three leaders as follows \cite{rastgoftar2016continuum}:
\begin{equation}
\label{alphaa}
\forall i\in \mathcal{V}_F,\qquad
\begin{bmatrix}
X_{1,0}&X_{2,0}&X_{3,0}\\
Y_{1,0}&Y_{2,0}&Y_{3,0}\\
1&1&1
\end{bmatrix}
    \begin{bmatrix}
    \alpha_{i,1}\\
    \alpha_{i,2}\\
    \alpha_{i,3}
    \end{bmatrix}
    =
    \begin{bmatrix}
    X_{i,0}\\
    Y_{i,0}\\
    1
    \end{bmatrix}
    .
\end{equation}

\section{Problem Statement}
\label{Problem Statement}
Consider a team of $N$ UAVs inside the leading triangle. Three leader UAVs, located at the vertices of the leading triangle, define the geometry of a triangle enclosing the follower UAVs. The paper makes the following assumptions:
\begin{enumerate}
    \item{Initial and target configurations of the leading triangle are known.}
    \item{The leading triangle must significantly deform to reach the target configuration.}
    \item{All UAVs have the same size and each UAV can be enclosed by a ball with radius $\epsilon>0$. }
\end{enumerate}

The objective is to minimize total UAV travel distance given initial and target configurations of the leading triangle such that risks to the overflown population and of collision are minimized. It is also desired that the team avoid flying over any "No-Fly-Zones" in the motion space. Let $\mathbf{r}=x\hat{\mathbf{e}}_1+y\hat{\mathbf{e}}_2\in \mathbf{R}^2$ and let $t$ denote time. Then,  $\mathcal{MO}=\mathcal{MO}\left(\mathbf{r}\right)\subset \mathbb{R}^2$ defines the motion space set and  $\Omega_{NFZ}(\mathbf{r)}\subset \mathcal{MO}$ defines the "No-Fly-Zone" in the motion space set.

We define the following legends:
\[
\begin{split}
    \mathcal{MO}\left(\mathbf{r}\right)\subset \mathbb{R}^2:&~\mathrm{motion}~\mathrm{space}\\
    \Omega_{NFZ}(\mathbf{r)}\subset \mathcal{MO}:&~\mathrm{No~Flight~Zone~Set} \\
    \Omega_{NAV}(\mathbf{r})=\mathcal{MO}\setminus \Omega_{NFZ} :&~\mathrm{Navigable~~Zone~Set}\\
\end{split}
\]
The above constrained optimization problem can be mathematically defined as follows:
\begin{equation}
    \min \sum_{i\in \mathcal{V}}\left(\zeta_{s,i}\int_{0}^{S_{F,i}}dS_i+\zeta_{h,i}\mathrm{Pr}\left(\mathrm{Human}|\mathbf{r}_i,t\right)\right)
\end{equation}
subject to rank condition \eqref{leaderrankk} and the following two conditions:
\begin{subequations}
\begin{equation}
\label{con2}
   \forall t\geq t_0, i_1,i_2\in \mathcal{V},i_1\neq i_2,\qquad \|\mathbf{r}_{i_1}-\mathbf{r}_{i_2}\|\geq 2\epsilon
\end{equation}
\begin{equation}
\label{con3}
   \forall t\geq t_0, \forall i\in \mathcal{V},\qquad \mathbf{r}_i\in \Omega_{NAV}.
\end{equation}
\end{subequations}
Note that $\zeta_{s,i}>0$ and $\zeta_{h,i}>0$ are constant scaling parameters and $\mathrm{Pr}\left(\mathrm{Human}|\mathbf{r}_i,t\right)$ assigns  likelihood of human presence on the navigable zone $\Omega_{NAV}$. Human presence probability, or population density, is treated as an optimization cost in this work. Furthermore,
\begin{equation}
    S_i=\int_{t_0}^t\sqrt{\left(\dfrac{\mathrm{d}x_{i,HT}}{\mathrm{d}t}\right)^2+\left(\dfrac{\mathrm{d}y_{i,HT}}{\mathrm{d}t}\right)^2}\mathrm{d}t
\end{equation}
is the path length of the UAV $i\in \mathcal{V}$. Also, $S_{F,i}$ is the length of  UAV $i$'s path connecting $\mathbf{R}_{i,0}$ and $\mathbf{R}_{i,F}$. Satisfaction of Eq. \eqref{leaderrankk} ensures that leaders form a convex hull at all times $t$. This is in fact a requirement for M{\color{black}U}S evolution as continuum deformation. The constraint Eq. \eqref{con2} ensures that no two UAVs approach closer than $2\epsilon$ in a continuum deformation coordination. Notice that inter-agent collision avoidance can be guaranteed if both conditions \eqref{leaderrankk} and \eqref{con2} are satisfied. In addition, condition \eqref{con3} ensures that the "No-Fly-Zone" is never entered by any UAV in the continuum deformation.

Assuming UAV $i\in \mathcal{V}$ moves on a straight path over $t\in [t_{k-1},t_k]$, we apply A* search to find the optimal path connecting initial and target positions of UAV $i\in \mathcal{V}$.

\section{Path-Planning}
\label{PathPlanning}
Suppose $\bar{\mathbf{T}}_c=(\mathbf{P}_{1,c},\mathbf{P}_{2,c},\mathbf{P}_{3,c})$ defines the desired configuration of the leading triangle in the $X-Y$ plane at the current time, where
\[
\begin{split}
    l=1,2,3,\qquad \mathbf{P}_{l,c}=&p_{x,l,c}\hat{\mathbf{e}}_1+p_{y,l,c}\hat{\mathbf{e}}_2\\
\end{split}
\]
is the position of leader $l\in \mathcal{V}_L$ expressed with respect to the ground coordinate system. The next desired configuration of the leading triangle is denoted by $\bar{\mathbf{T}}_{n}=(\mathbf{P}_{1,n},\mathbf{P}_{2,n},\mathbf{P}_{3,n})$, where
\[
\begin{split}
    l=1,2,3,\qquad \mathbf{P}_{l,n}=&p_{x,l,n}\hat{\mathbf{e}}_1+p_{y,l,n}\hat{\mathbf{e}}_2.\\
\end{split}
\]
Leaders' waypoints are obtained by uniform discretization of the $X-Y$ plane, where
\begin{equation}
\label{pln1}
    q=x,y,\qquad p_{q,l,n}=p_{q,l,c}+h_q\Delta p_q
\end{equation}
and 
\begin{equation}
\label{pln2}
    q=x,y,\qquad h_q\in \{-1,0,1\}.
\end{equation}

The paper assumes initial and target configurations of the leading triangle are given. 
In addition, $\bar{\mathbf{T}}_g=(\mathbf{P}_{1,g},\mathbf{P}_{2,g},\mathbf{P}_{3,g})$ and $\bar{\mathbf{T}}_0=(\mathbf{P}_{1,0},\mathbf{P}_{2,0},\mathbf{P}_{3,0})$ are the goal and initial configurations of the leading triangle $j\in \Omega_{CL}$, where
\[
\begin{split}
    l=1,2,3,\qquad \mathbf{P}_{l,g}=&p_{x,l,g}\hat{\mathbf{e}}_1+p_{y,l,g}\hat{\mathbf{e}}_2\\
    l=1,2,3,\qquad \mathbf{P}_{l,0}=&p_{x,l,0}\hat{\mathbf{e}}_1+p_{y,l,0}\hat{\mathbf{e}}_2\\
\end{split}
.
 \]

Assuming leaders  move on a straight path, the path of leader $l\in \mathcal{V}_L$  is defined by
\begin{equation}
\label{rlht}
\begin{split}
   l\in \mathcal{V}_L,\qquad\mathbf{r}_{l,HT}=\left(1-\beta\right)\mathbf{P}_{l,c}+\beta\mathbf{P}_{l,n}+z_{HT}\hat{\mathbf{e}}_3,
\end{split}
\end{equation}
where MQS elevation $z_{HT}$ is constant,  $l\in \mathcal{V}_{L}^{j}, ~j\in \Omega_{CL}$, and  $\beta\in [0,1]$.

\begin{theorem}\label{theorem1}
Let $d_s$ be the minimum separation distance of two UAVs at initial time $t_0$, $d_b$ be the minimum distance of a UAV from the sides of the leading triangle at time $t_0$, and each UAV be enclosed by a ball with radius $\epsilon$. Define
\begin{equation}
\label{deltamax}
    \delta_{max}=\mathrm{min}\bigg\{{1\over 2}\left(d_s-2\epsilon\right),\left(d_b-\epsilon\right)\bigg\}.
\end{equation}
Let $\mathbf{r}_{i,HT}$  be the global desired position of UAV $i\in \mathcal{V}$, given by a continuum deformation (See Eq. \eqref{maincontinum}), $\mathbf{r}_i$ be the be the actual position of UAV $i\in \mathcal{V}$, and $\delta$ be the  upper limit for deviation of UAV $i\in \mathcal{V}$ from continuum deformation desired position:
\begin{equation}
   t\in [t_k,t_{k+1}],~ \forall i\in \mathcal{V},\qquad \|\mathbf{r}_i-\mathbf{r}_{i,HT}\|\leq \delta.
\end{equation}
Define 
\begin{equation}
    \lambda_{\mathrm{CD,min}}=\dfrac{\delta+\epsilon}{\delta_{max}+\epsilon}.
\end{equation}
If 
\begin{equation}
\label{ccolk}
   t\in [t_k,t_{k+1}],\qquad \mathcal{C}_{Col,k}=\lambda_{\mathrm{CD,min}}- \lambda_1\left(U_{\mathrm{CD}}\right)\leq 0,
\end{equation}
then,
\begin{enumerate}
    \item{Inter-agent collision avoidance is guaranteed and}
\item{All followers remain inside the leading triangle $j$ at any $\beta\in [0,1]$.}
\end{enumerate}
\end{theorem}
\textbf{Proof:} See the proof in \cite{rastgoftar2016asymptotic}.

\textbf{Corollary:} If the constraint Eq. \eqref{ccolk} is met, then, we can guarantee that no two UAVs collide (i.e., Eq. \eqref{con2} is satisfied).

\textbf{Definition (Valid Continuum Deformation)}: A leading triangle configuration $\bar{\mathbf{T}}_n$ is called a \textit{valid deformation}, if 
\begin{enumerate}
    \item{$\mathbf{P}_{l,n}$ is defined by Eqs. \eqref{pln1} and \eqref{pln2} and}
    \item{Constraint Eqs. \eqref{leaderrankk}, \eqref{con2}, and \eqref{con3} are all satisfied.}
\end{enumerate}

\subsection{Continuum Deformation Optimization}
\label{Continuum Deformation Motion Planning}
This paper applies A* search to optimally plan the continuum deformation via its leading triangle.  We define the following legends:
\[
\begin{split}
    s_0=\left(\bar{\mathbf{T}}_0,t_0\right):=&\mathrm{Initial~node}\\
    s_g=\left(\bar{\mathbf{T}}_g,t_{g}\right):=&\mathrm{Goal~node}\\
    s_c=\left(\bar{\mathbf{T}}_c,t_c\right):=&\mathrm{Current~node}\\
    s_{n}=\left(\bar{\mathbf{T}}_{n}^1,t_n\right):=&\mathrm{Next~node}\\
\end{split}
\]
where $\Delta t=t_n-t_c$ is time increment. Leaders' optimal paths are determined by minimizing continuum deformation cost given by
\begin{equation}
    F\left(s_n\right)=G\left(s_n\right)+H\left(s_n\right),
\end{equation}
where $s_n$ is a valid continuum deformation and $h\left(s_n\right)$ is the heuristic cost assigned as follows:
\begin{equation}
    H\left(s_n\right)=\sqrt{\sum_{l\in \mathcal{V}_L}\bigg\|\mathbf{P}_{l,n}-\mathbf{P}_{l,g}\bigg\|^2},
\end{equation}
Furthermore, $g\left(s_n\right)$ is the minimum estimated cost from $s_0$ to $s_n$
\begin{equation}
    G\left(s_n\right)=\min\big\{G\left(s_{c}\right)+C_{c,n}\big\},
\end{equation}
where
\begin{equation}
\begin{split}
    C_{c,n}=&\sum_{l\in \mathcal{V}_L}\zeta_{s,l}\bigg\|\mathbf{P}_{l,n}^{j}-\mathbf{P}_{l,c}^{j}\bigg\|\\
    +&\sum_{l\in \mathcal{V}_L}\zeta_{h,l}\bigg|\mathrm{Pr}\left(\mathrm{Human}|\mathbf{P}_{l,n},t_n\right)-\mathrm{Pr}\left(\mathrm{Human}|\mathbf{P}_{l,c},t_c\right)\bigg|
    .
\end{split}
\end{equation}


\section{Trajectory Planning}
\label{Continuum Deformation Acquisition}
\subsection{Leaders' Desired Trajectories}
Leaders' paths are all prescribed as piece-wise linear. To ensure that leaders' trajectories are $\mathcal{C}^2$ continuous, $\beta$ in Eq. \eqref{rlht} is given by a fifth order polynomial:
\begin{equation}
\label{poly1}
    t\in [t_{k-1},t_k], \qquad \beta\left(t\right)=\sum_{i=0}^5=a_{i,k}t^{5-i}
\end{equation}
subject to 
\begin{equation}
\label{poly2}
\begin{split}
    \beta\left(t_{k-1}\right)=&1\\
    \beta\left(t_k\right)=&0\\
    \dot{\beta}\left(t_{k-1}\right)=\dot{\beta}\left(t_{k}\right)&=0\\
    \ddot{\beta}\left(t_{k-1}\right)=\ddot{\beta}\left(t_{k}\right)&=0.\\
\end{split}
\end{equation}
Assuming $\Delta t=t_k-t_{k-1}$ ($\forall k$), $a_{0,k}$ through $a_{5,k}$ are determined by solving the following linear equality constraints:
\begin{equation}
    \begin{bmatrix}
    0&0&0&0&0&1\\
    \Delta t^5&\Delta t^4&\Delta t^3&\Delta t^2&\Delta t&1\\
    0&0&0&0&1&0\\
    5\Delta t^4&4\Delta t^3&3\Delta t^2&2\Delta t&1&0\\
     0&0&0&2&0&0\\
    20\Delta t^3&12\Delta t^2&6\Delta t&2&0&0\\
    \end{bmatrix}
    \begin{bmatrix}
    a_{0,k}\\
    a_{1,k}\\
    a_{2,k}\\
    a_{3,k}\\
    a_{4,k}\\
    a_{5,k}
    \end{bmatrix}
    =
    \begin{bmatrix}
    0\\
    1\\
    0\\
    0\\
    0\\
    0
    \end{bmatrix}
    .
\end{equation}

\subsection{Followers' Desired Trajectories}
\label{Desired Coordination}
Follower $i$'s desired position is a convex combination of leaders desired positions as defined in Eq. \eqref{globdespos}.  Substituting $\mathbf{r}_{l,HT}$ by Eq. \eqref{rlht}, desired position of follower UAV $i$ is expressed as follows:
\begin{equation}
\begin{split}
    i\in \mathcal{V}_F,\qquad \mathbf{r}_{i,HT}=\sum_{l\in \mathcal{V}_L}\alpha_{i,l}\bigg[\left(1-\beta\right)\mathbf{P}_{l,c}+\beta\mathbf{P}_{l,n}\bigg]+z_{HT}\hat{\mathbf{e}}_3
\end{split}
\end{equation}
 \section{UAV Dynamics}
 \label{UAV Dynamics}
 Dynamics of UAV $i\in \mathcal{V}_F$ is given by
\begin{equation}
\label{UAVdynamics1}
   \begin{split}
   \dot{\mathbf{r}}_i=&\mathbf{v}_i\\
   \dot{\mathbf{v}}_i=&
   \begin{bmatrix}
0&
0&
-g
\end{bmatrix}
^T
+
\bar{F}_{T,i}
\hat{\mathbf{k}}_{b,i}\\
   \begin{bmatrix}
   \ddot{\bar{F}}_{T,i}&
   \ddot{\mathbf{\phi}}_i&
   \ddot{\mathbf{\theta}}_i&
   \ddot{\mathbf{\psi}}_i
   \end{bmatrix}
   ^T
   =&
   \begin{bmatrix}
   u_{T,i}&{u}_{\phi,i}&{u}_{\theta,i}&u_{\psi,i}
   \end{bmatrix}
   ^T
 \end{split}
\end{equation}
Note that $\phi_i$, $\theta_i$, and $\psi_i$ are UAV $i$'s Euler angles, $m_i$ is mass, thrust force $F_{T,i}$$g=9.81 {m\over {s^2}}$ is the gravity, $\bar{F}_{T,i}={F_{T,i}\over m_i}$ is thrust force per mass $m_i$, and $\hat{\mathbf{k}}_{b,i}$ is the unit vector assigning direction of the thrust force $\bar{F}_{T,i}$. 


Dynamics \eqref{UAVdynamics1} can be rewritten in the following form: 
\begin{equation}
\label{UAVdynamics}
\begin{cases}
    \dot{\mathcal{X}}_i=\mathbf F_i\left(\mathcal{X}_i\right)+G\mathcal{V}_i\\
    \mathbf{r}_i=h_i\left(\mathcal{X}_i\right)=[x_i~y_i~z_i]^T\\
\end{cases}
\end{equation}
where
\[
 \mathcal{X}_i=[x_i~y_i~z_i~v_{x,i}~v_{y,i}~v_{z,i}~\bar{F}_{T,i}~\phi_i~\theta_i~\psi_i~\dot{\bar{F}}_{T,i}~\dot{\phi}_i~\dot{\theta}_i~\psi_i]^T 
\]
is the control state, $\mathbf{r}_i$ is the control output, and
$\mathbf{V}_i=[u_{T,i}~u_{\phi,i}~u_{\theta,i}]$ is the control input vector. 
\begin{subequations}
\begin{equation}
\begin{split}
     \mathbf{F}_i=&[v_{x,i}~v_{y,i}~v_{z,i}~f_{4,i}~f_{5,i}~f_{6,i}~\dot{\bar{F}}_{T,i}~\dot{\phi}_i~\dot{\theta}_i~\dot{\psi}_i~0~0~0~f_{14,i}]^T\\
     \begin{bmatrix}
         f_{4,i}\\
         f_{5,i}\\
         f_{6,i}\\
     \end{bmatrix}=&
     \begin{bmatrix}
        0\\
        0\\
        -g
        \end{bmatrix}
        +
        \bar{F}_{T,i}
    \begin{bmatrix}
        C_{\phi_{i}}S_{\theta_{i}}C_{\psi_{i}}+S_{\phi_{i}}S_{\psi_{i}}\\
        C_{\phi_{i}}S_{\theta_{i}}S_{\psi_{i}}-S_{\phi_{i}}C_{\psi_{i}}\\
        C_{\phi_{i}}C_{\theta_{i}}\\
    \end{bmatrix}
\end{split}
,
\end{equation}
\begin{equation}
    G=
    \begin{bmatrix}
        0_{9\times 3}\\
        I_3\\
        0_{1,3}\\
    \end{bmatrix}
    .
\end{equation}
\end{subequations}
\textbf{Yaw Control}: In this paper, $\ddot{\psi}_i=u_{\psi,i}$ is chosen as follows:
\[
u_{\psi,i}=\ddot{\psi}_{d,i}+k_{\dot{\psi}_i}\left(\dot{\psi}_{d,i}-\ddot{\psi}_{i}\right)+k_{{\psi}_i}\left(\dot{\psi}_{d,i}-\ddot{\psi}_{i}\right),
\]
where $k_{\psi_i}>0$ and $k_{\dot{\psi}_i}>0$ are constant. It is assumed that $\psi_{d,i}$, $\dot{\psi}_{d,i}$, and $\ddot{\psi}_{d,i}$ are known. Therefore, 
$\psi_i$ is updated as follows:
\begin{equation}
    \left(\ddot{\psi}_{i}-\ddot{\psi}_{d,i}\right)+k_{\dot{\psi}_i}\left(\dot{\psi}_{i}-\dot{\psi}_{d,i}\right)+k_{{\psi}_i}\left({\psi}_{i}-{\psi}_{d,i}\right)=0.
\end{equation} 

 \section{UAV Control}
 \label{UAV Control}
 \subsection{Outer-Loop Control}
 Desired dynamics of UAV $i$ is given by
 \begin{subequations}
 \begin{equation}
 i\in \mathcal{V},\qquad     \ddot{\mathbf{r}}_i=\mathbf{U}_i
 \end{equation}
 \begin{equation}
 i\in \mathcal{V},\qquad \mathbf{U}_i=\mathbf{L}_{d,i}-\mathbf{L}_i,
 \end{equation}
 \begin{equation}
 i\in \mathcal{V},\qquad \mathbf{L}_{d,i}=\ddot{\mathbf{r}}_{i,HT}+\gamma_{1,i}\dot{\mathbf{r}}_{i,HT}+\gamma_{2,i}{\mathbf{r}}_{i,HT},
 \end{equation}
 \begin{equation}
 i\in \mathcal{V},\qquad \mathbf{L}_i=\gamma_{1,i}\dot{\mathbf{r}}_{i}+\gamma_{2,i}{\mathbf{r}}_{i},
 \end{equation}
 \end{subequations}
 where $\gamma_{1,i}>0$ and $\gamma_{2,i}>0$ are constant. Therefore, dynamics of every UAV $i\in \mathcal{V}$ is stable. In addition, $\mathbf{U}_i=[u_{1,i}~u_{2,i}~u_{3,i}]^T$ is a fictitious input used to determine desired thrust $\bar{F}_{T,i}$, $\phi_{T,i}$, $\theta_{d,i}$:
 
\begin{subequations}
\begin{equation}
    \bar{{F}}_{T,d,i}=\|\mathbf{U}_i\|,
\end{equation}
\begin{equation}
    \phi_{d,i} = -\sin^{-1} \left( \dfrac{u_{1,i}S_{\psi_i} - u_{2,i}C_{\psi_i}}{\|\mathbf{U}_i\|} \right ),
\end{equation}
\begin{equation}
    \theta_{d,i} = \tan^{-1} \left( \dfrac{u_{1,i}C_{\psi_i} + u_{2,i}S_{\psi_i}}{u_{3,i}} \right ).
\end{equation}
\end{subequations}
 \subsection{{\color{black}Inner}-Loop Control}
\begin{figure}
\center
\includegraphics[width=6 in]{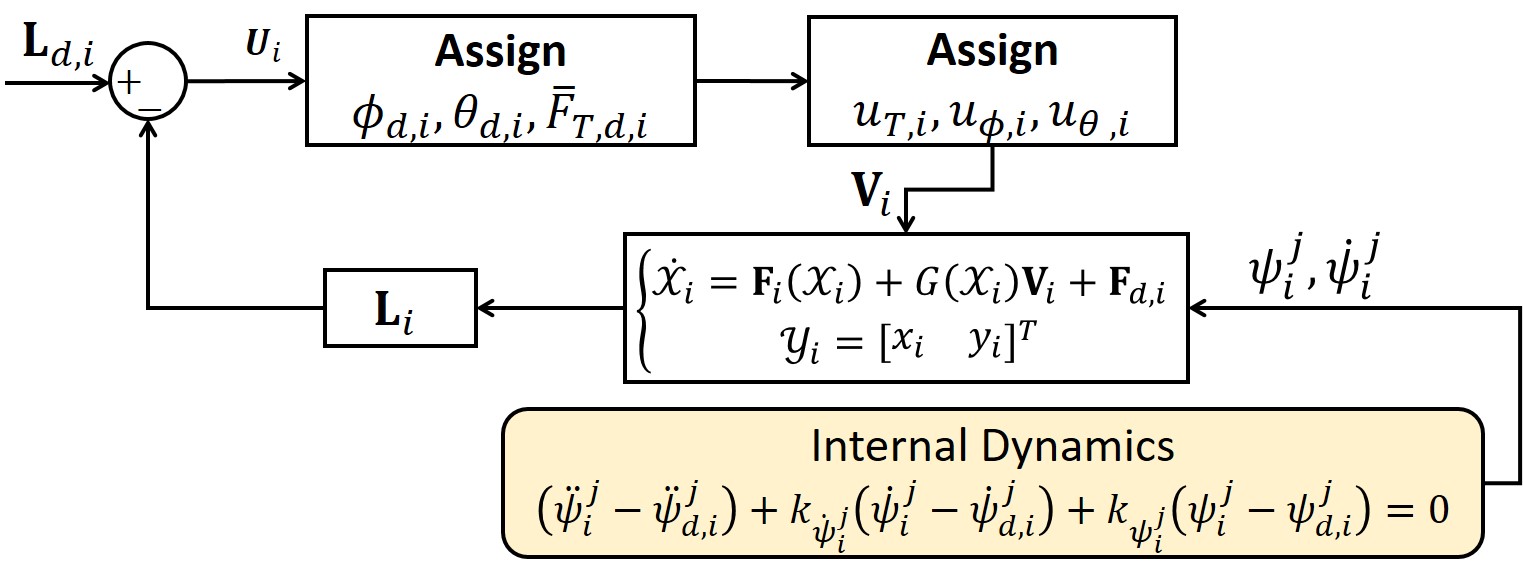}
\caption{UAV controller block diagram.}
\label{controllerblockdiagram}
\end{figure}

The UAV $i$ control input $\mathbf{V}_i=[u_{T,i}~u_{\phi,i}~u_{\theta,i}]^T$ is chosen as follows:
\begin{equation}
    \begin{bmatrix}
    u_{T,i}\\
    u_{\theta,i}\\
    u_{\psi,i}
    \end{bmatrix}
    =
    \begin{bmatrix}
    -k_{\dot{T}_i}\dot{\bar{F}}_{T,i}+k_{{T}_i}\left({\bar{F}}_{T,d,i}-{\bar{F}}_{T,i}\right)\\
    -k_{\dot{\phi}_i}\dot{{\phi}}_{i}+k_{{\phi}_i}\left({\phi}_{d,i}-{\phi}_{i}\right)\\
    -k_{\dot{\theta}_i}\dot{{\theta}}_{i}+k_{{\theta}_i}\left({\theta}_{d,i}-{\theta}_{i}\right)\\
    \end{bmatrix}
   .
\end{equation}
The block digram of UAV $i\in \mathcal{V}$ controller is shown in Fig. \ref{controllerblockdiagram}.
\begin{figure*}
 \centering
  \subfigure[]{\includegraphics[width=0.23\linewidth]{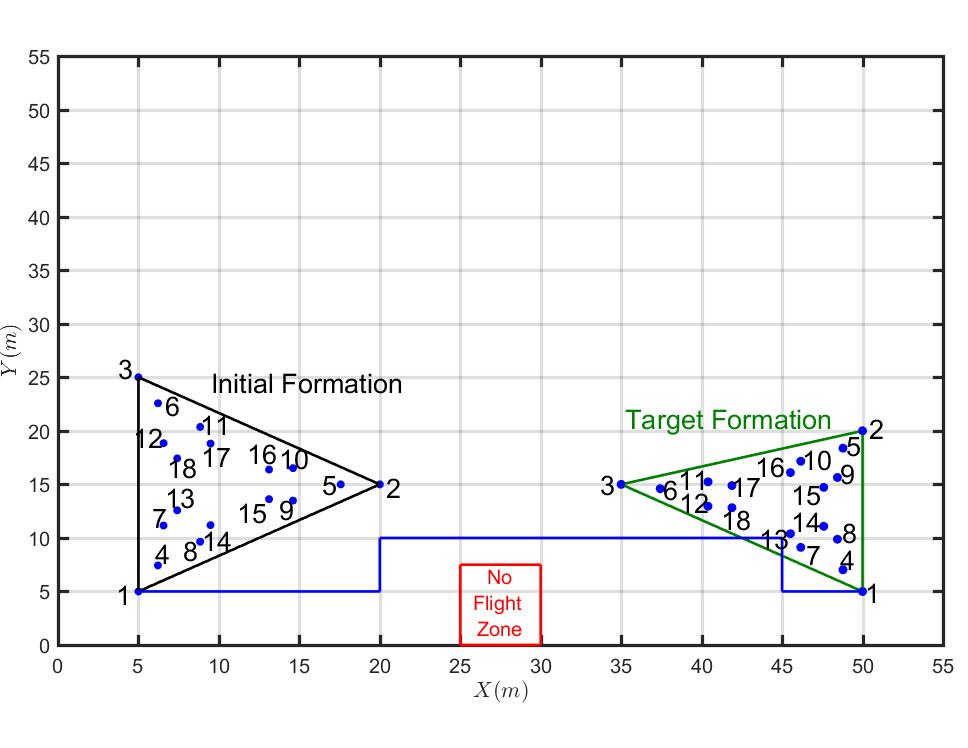}}
  \subfigure[]{\includegraphics[width=0.23\linewidth]{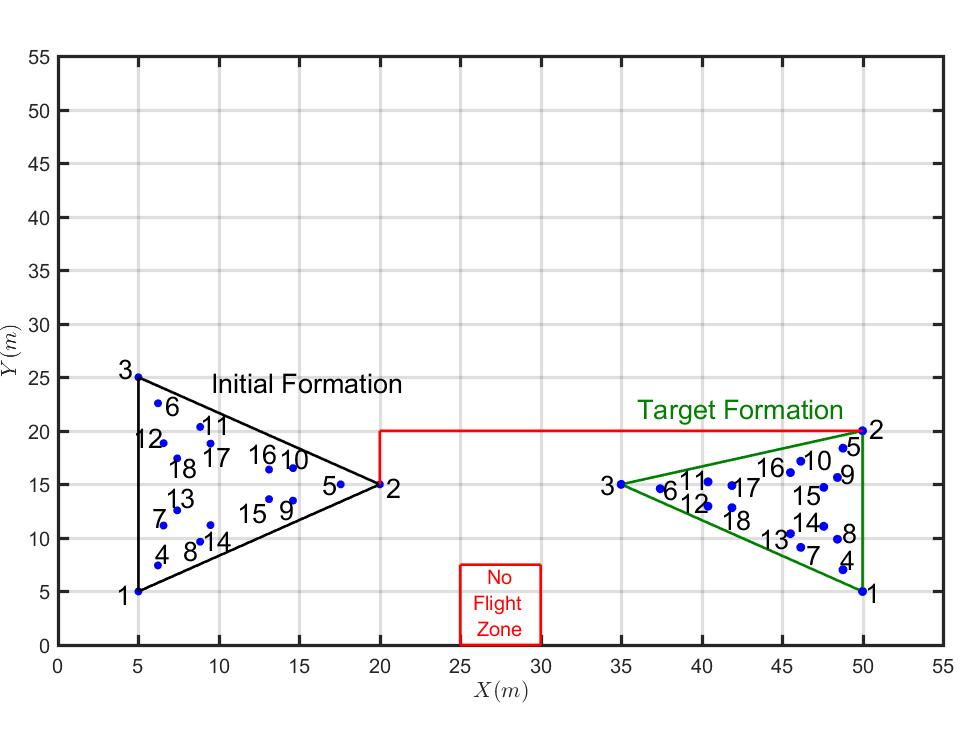}}
  \subfigure[]{\includegraphics[width=0.23\linewidth]{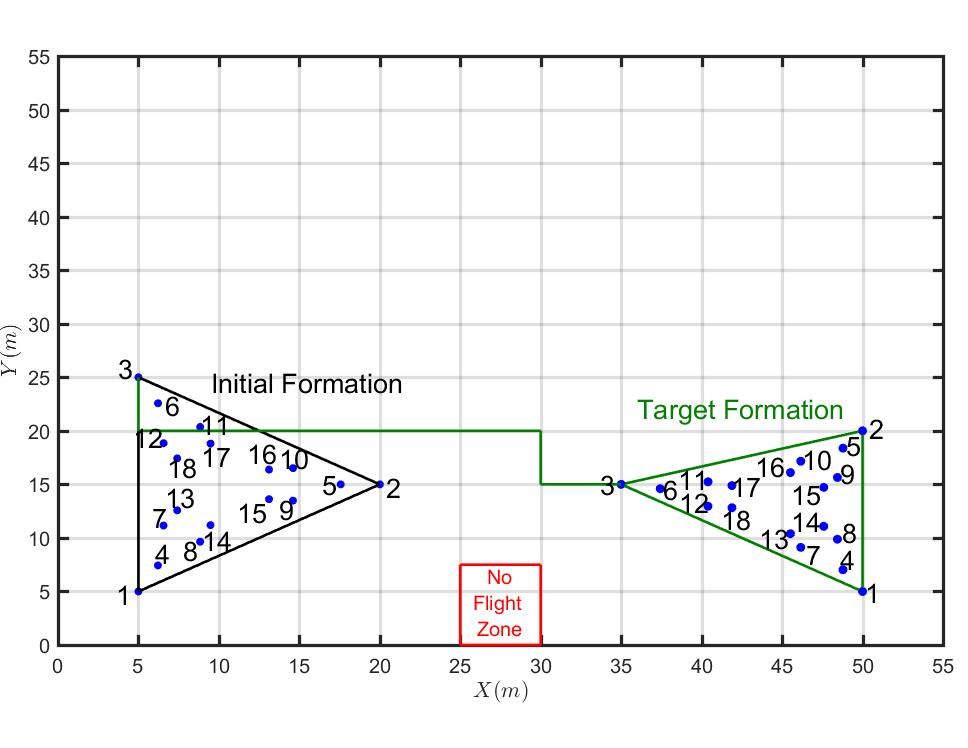}}
  \subfigure[]{\includegraphics[width=0.23\linewidth]{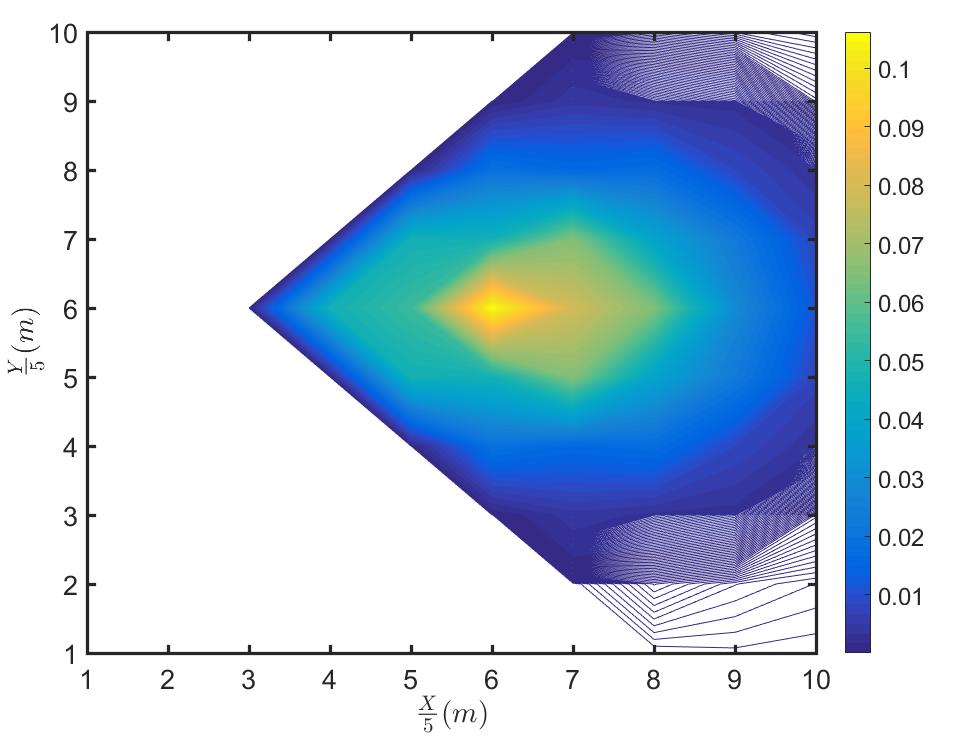}}
  \caption{(a-c) Leaders' optimal paths, MQS initial configuration, and MQS target formation in case study 1 ((a) Path of leader $2$, (b) Path of leader $2$, and (c) Path of leader $3$). (d) Probability distribution contour assigning likelihood of human presence $\mathrm{Pr}\left(\mathrm{Human}|\mathbf{r}_i\right)$. {\color{black}No human is present in the no-color area.}}
\label{case1}
\end{figure*}

\section{Simulation Results}
\label{Simulation Results}
In this section, we simulate continuum deformation of an MQS in a complex environment. We consider two cases. In the first case-study, MQS continuum deformation is planned given human presence probability distribution over the motion field. In the second case study, MQS continnum deformation is optimized given deterministic motion of a human.
\subsection{Case Study 1}
Consider a MQS consisting of $18$ UAVs with  initial formation shown in Fig. \ref{case1} (a-c). Leaders are initially positioned at 
$\mathbf{R}_{1,0}=5\hat{\mathbf{e}}_1+5\hat{\mathbf{e}}_2+10\hat{\mathbf{e}}_3$, $\mathbf{R}_{2,0}=20\hat{\mathbf{e}}_1+15\hat{\mathbf{e}}_2+10\hat{\mathbf{e}}_3$,
 $\mathbf{R}_{3,0}=5\hat{\mathbf{e}}_1+25\hat{\mathbf{e}}_2+10\hat{\mathbf{e}}_3$. It is desired that leaders ultimately form the triangular formation shown in Fig.  \ref{case1}. Leaders' target destinations are $\mathbf{R}_{1,F}=50\hat{\mathbf{e}}_1+5\hat{\mathbf{e}}_2+10\hat{\mathbf{e}}_3$, $\mathbf{R}_{2,F}=50\hat{\mathbf{e}}_1+20\hat{\mathbf{e}}_2+10\hat{\mathbf{e}}_3$,
, $\mathbf{R}_{3,F}=35\hat{\mathbf{e}}_1+15\hat{\mathbf{e}}_2+10\hat{\mathbf{e}}_3$. Note that the MQS needs to significantly deform and rotate in order to reach the target formation from the initial configuration shown in Fig.  \ref{case1}. Furthermore, the MUS must avoid flying over the "No-Fly-Zone" shown by the red box in Fig.   \ref{case1}. Additionally, it is preferable that the MQS flies over unpopulated or sparsely-populated areas. Therefore, the likelihood of human presence, e.g., based on census data, is considered as navigation cost.
It is assumed that the likelihood of human presence is time invariant ($\mathrm{Pr}\left(\mathrm{Human}|\mathbf{r}_i,t\right)=\mathrm{Pr}\left(\mathrm{Human}|\mathbf{r}_i\right),\forall t$) as shown in Fig. \ref{case1} (d).
\begin{figure}
\center
\includegraphics[width=6 in]{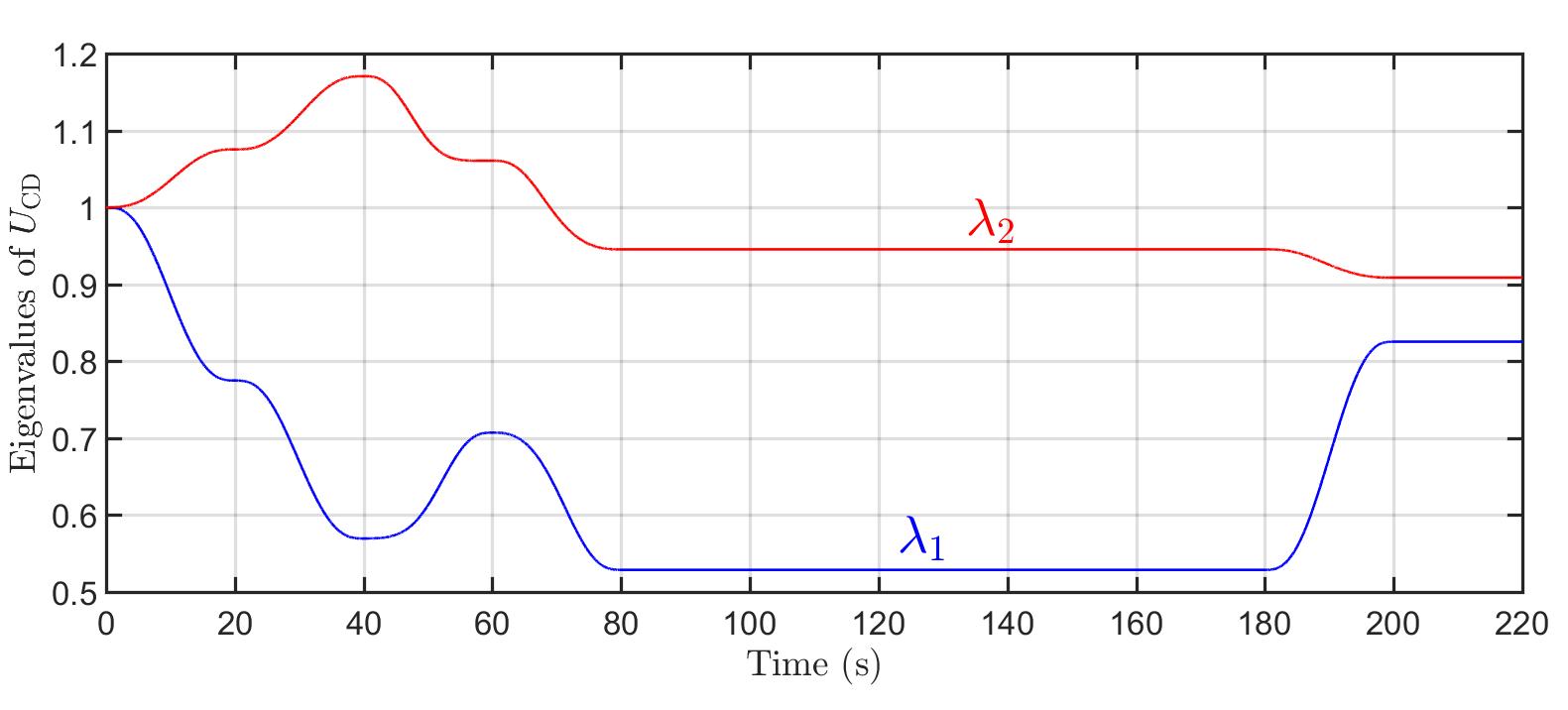}
\caption{Eigenvalues of matrix $U_{\mathrm{CD}}$ versus time in case-study $1$.}
\label{eigenvaluesudprobabilistic}
\end{figure}
\begin{figure}
\center
\includegraphics[width=6 in]{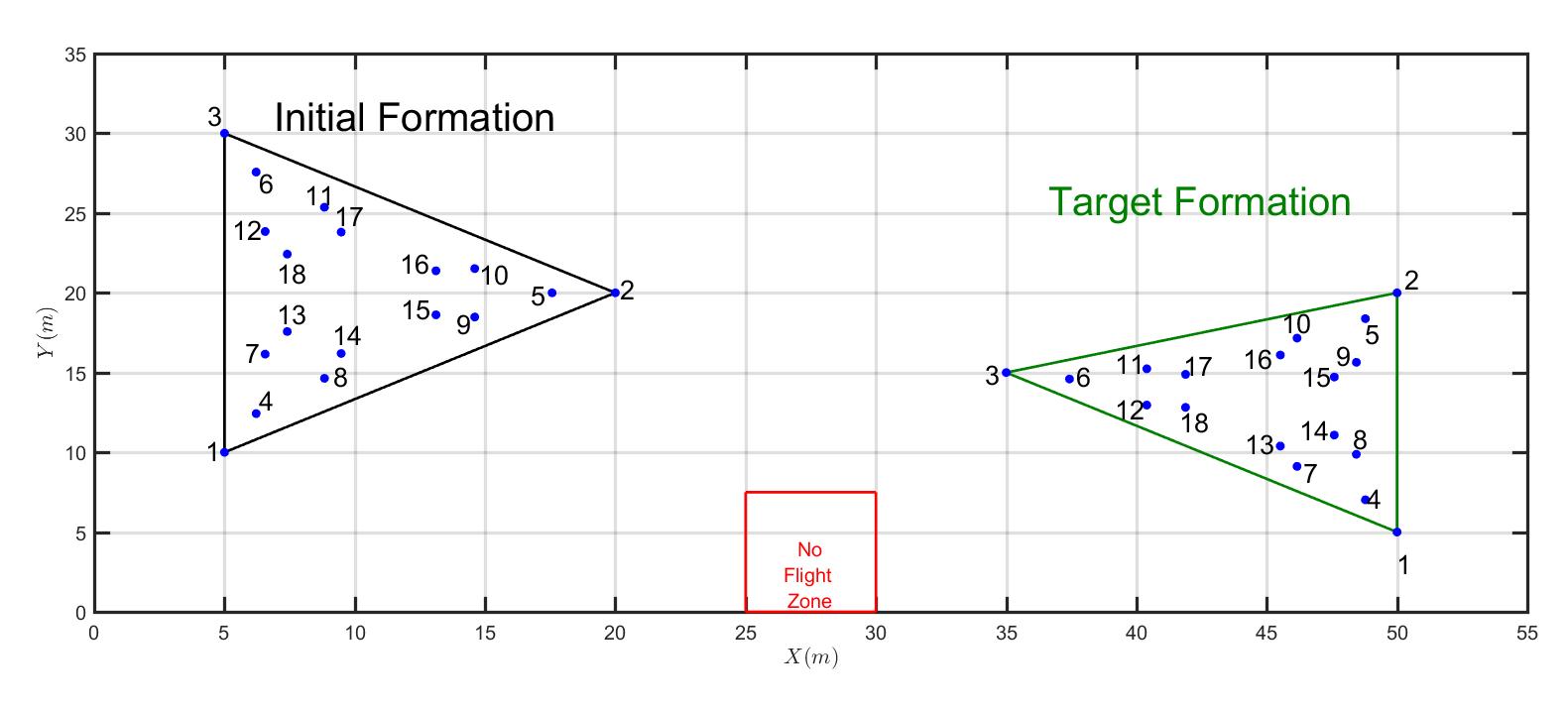}
\caption{MQS initial and target formations in case study 2.}
\label{case2initfinal}
\end{figure}
Given leaders' optimal trajectories, eigenvalues of pure deformation matrix $U_{\mathrm{CD}}$ are plotted versus time in Fig.  \ref{eigenvaluesudprobabilistic}.

\begin{figure}
 \centering
  \subfigure[]{\includegraphics[width=0.8\linewidth]{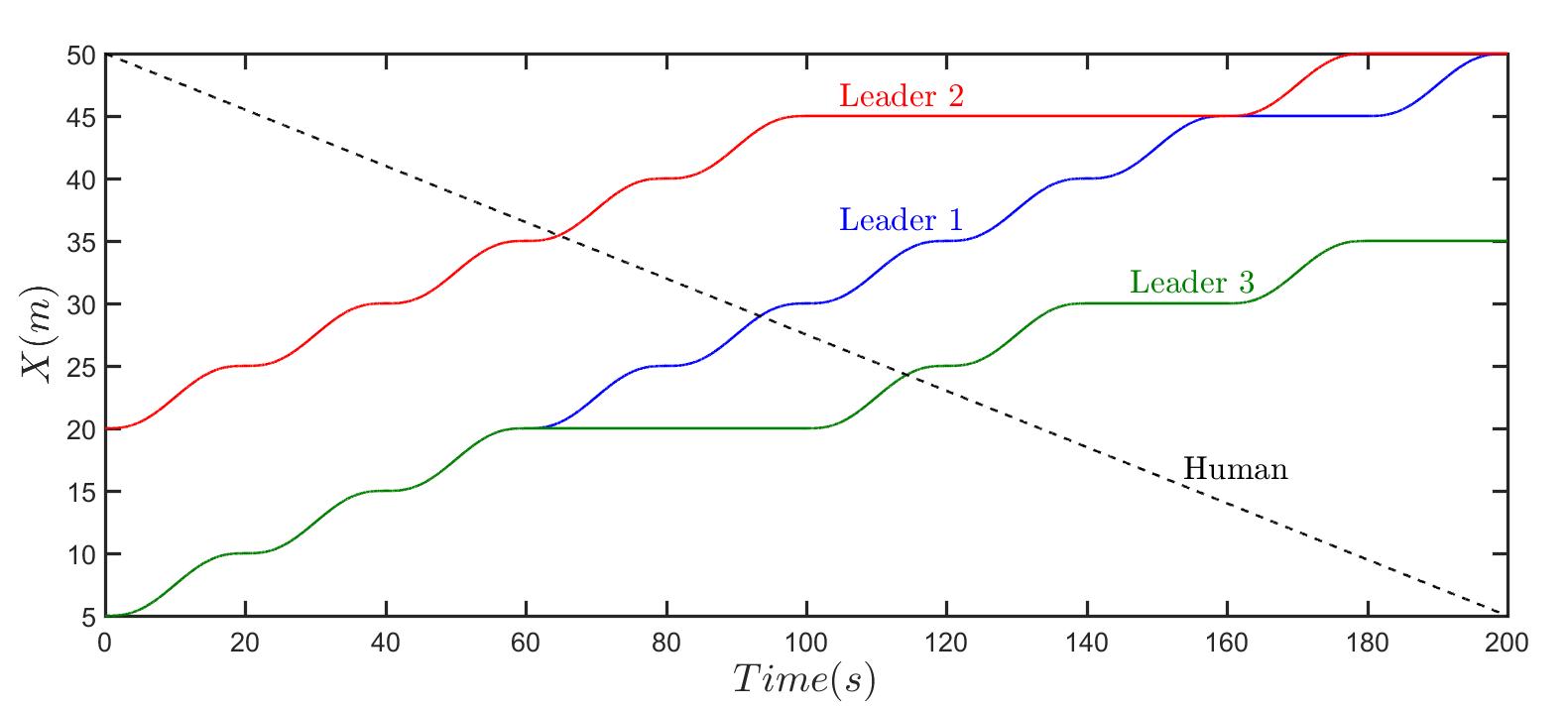}}
  \subfigure[]{\includegraphics[width=0.8\linewidth]{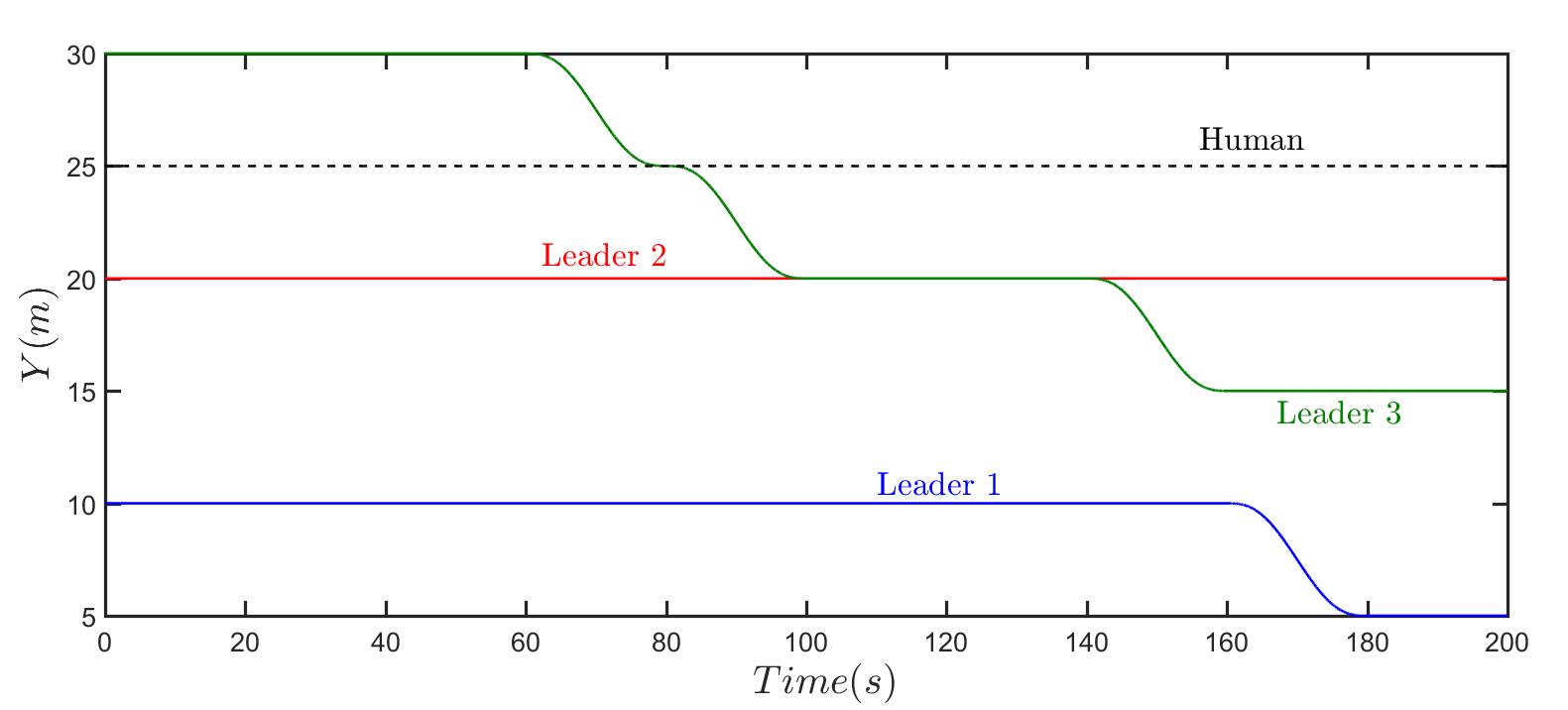}}
  \caption{(a,b) $X$ and $Y$ components of leaders' optimal trajectories and the simulated human trajectory.}
\label{leadeshumancase2}
\end{figure}

\begin{figure}
\center
\includegraphics[width=6 in]{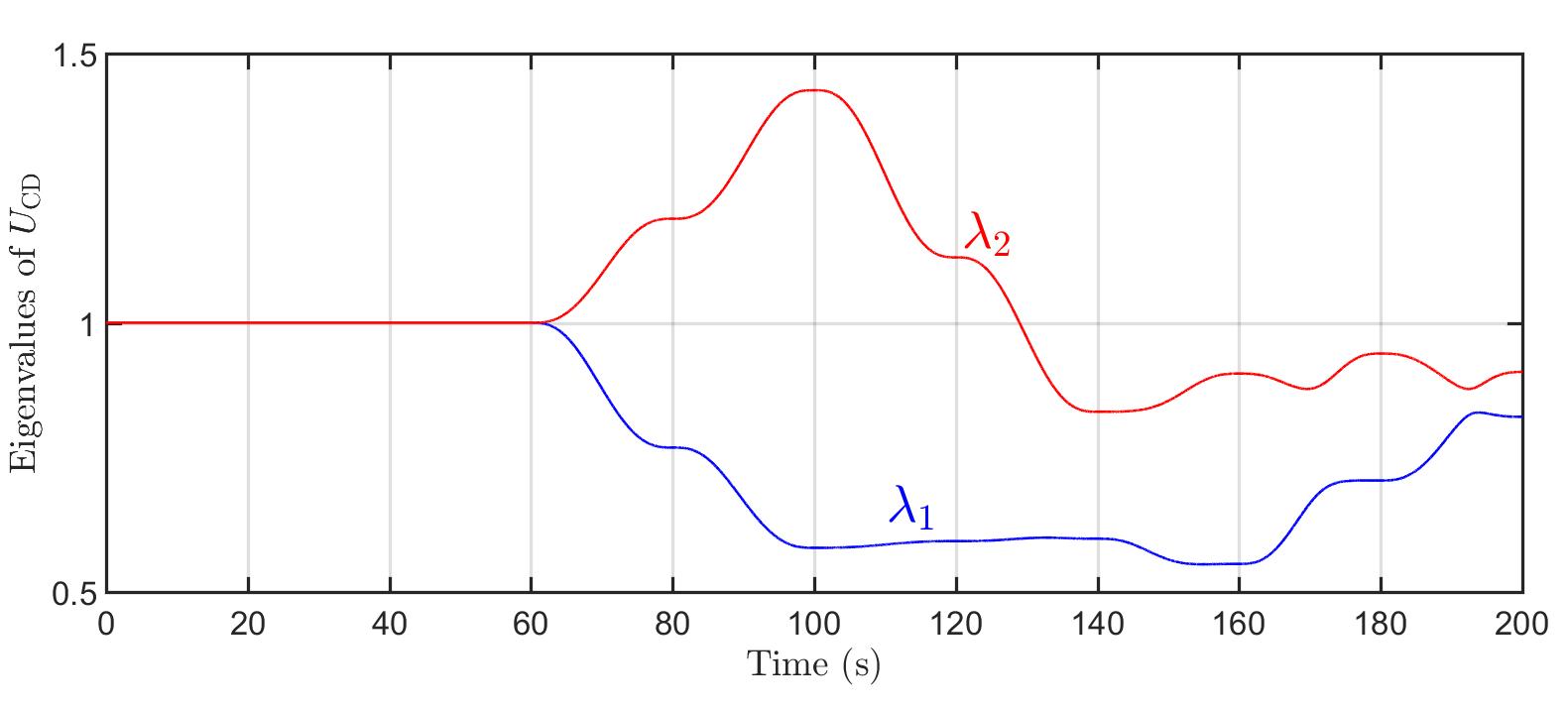}
\caption{Eigenvalues of matrix $U_{\mathrm{CD}}$ versus time in case-study $2$.}
\label{eigenvaluesuddeterministic}
\end{figure}

\begin{figure}
 \centering
  \subfigure[$t=0s$]{\includegraphics[width=0.3\linewidth]{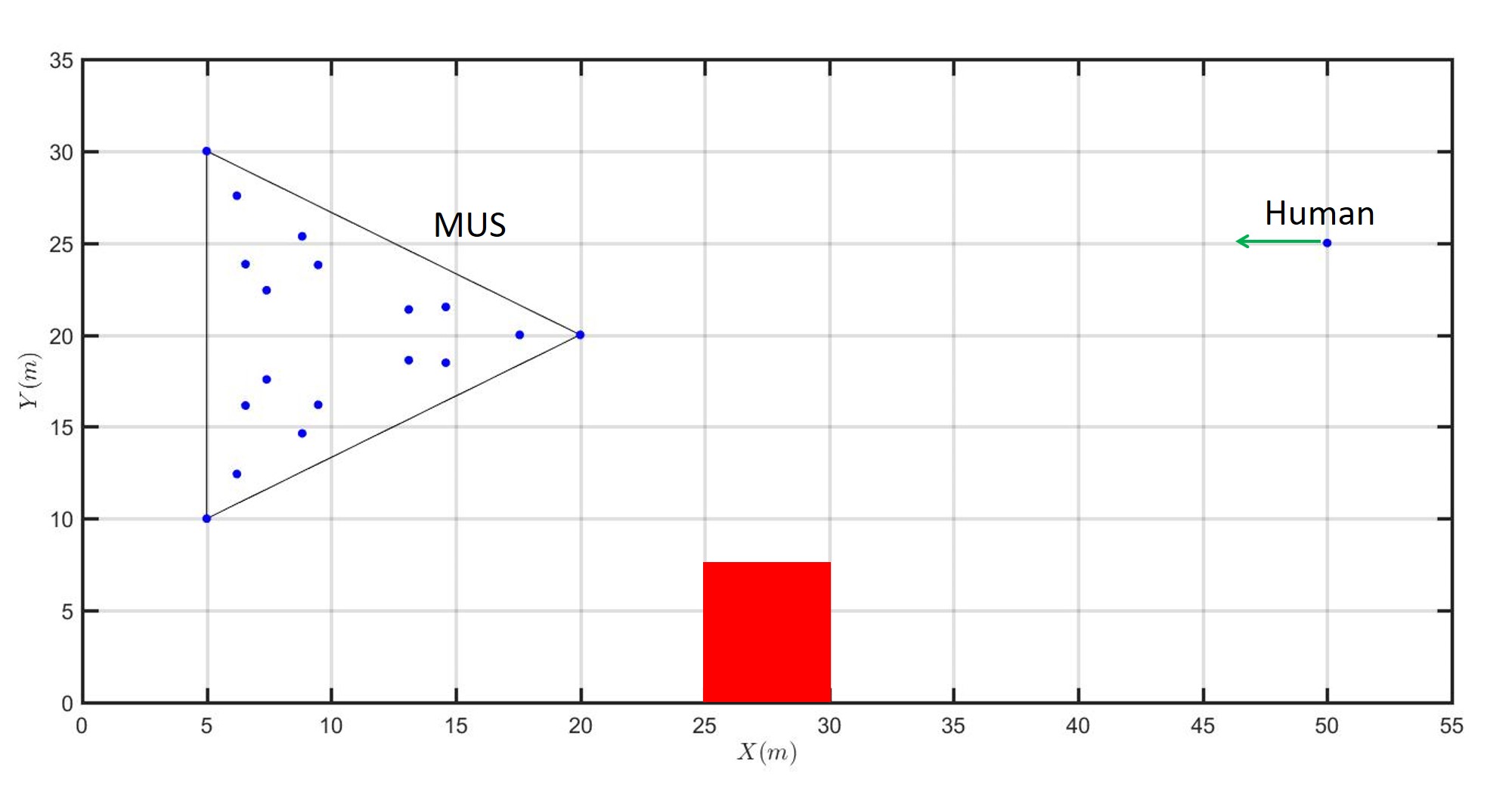}}
  \subfigure[$t=25s$]{\includegraphics[width=0.3\linewidth]{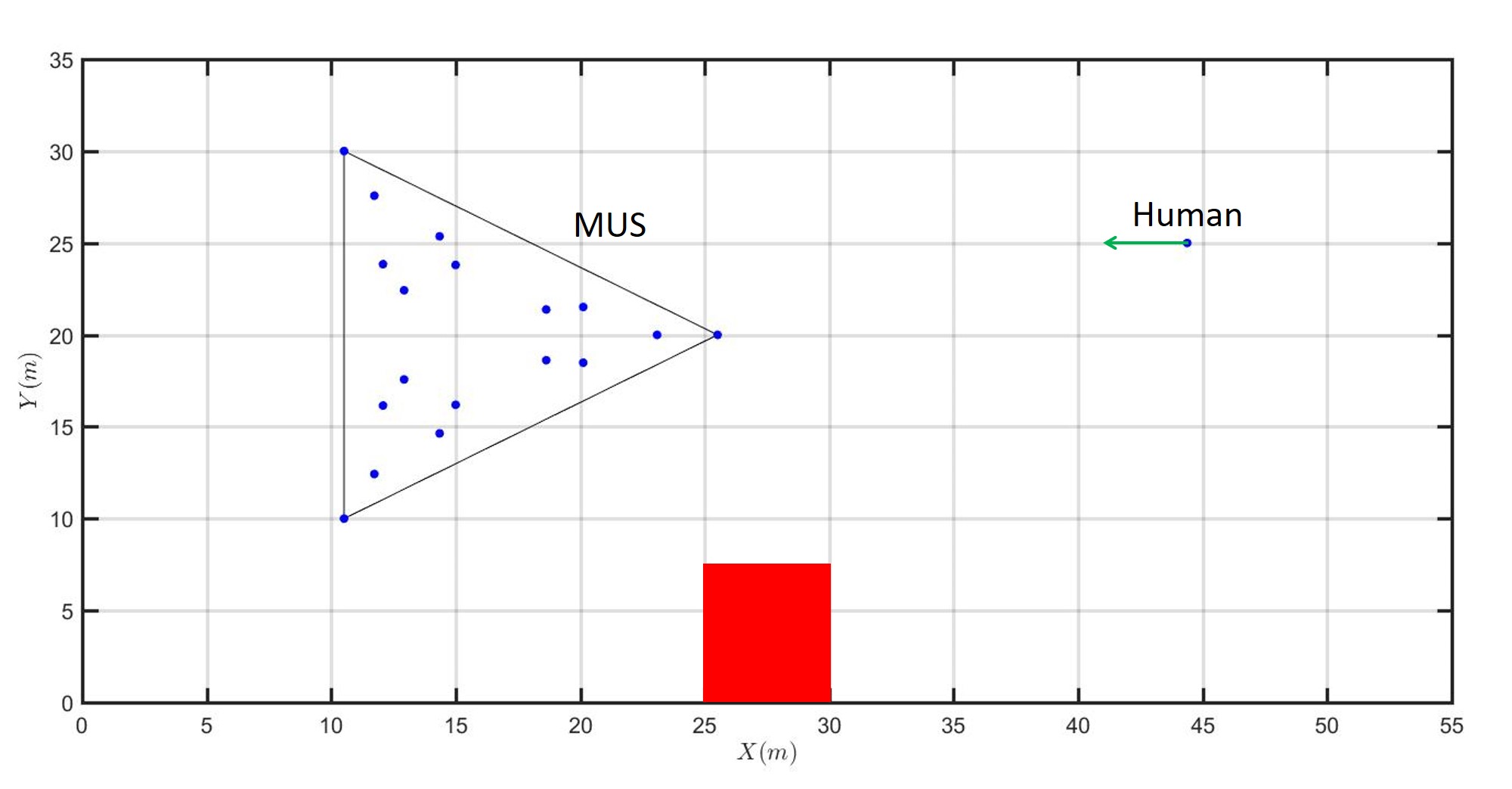}}
  \subfigure[$t=50s$]{\includegraphics[width=0.3\linewidth]{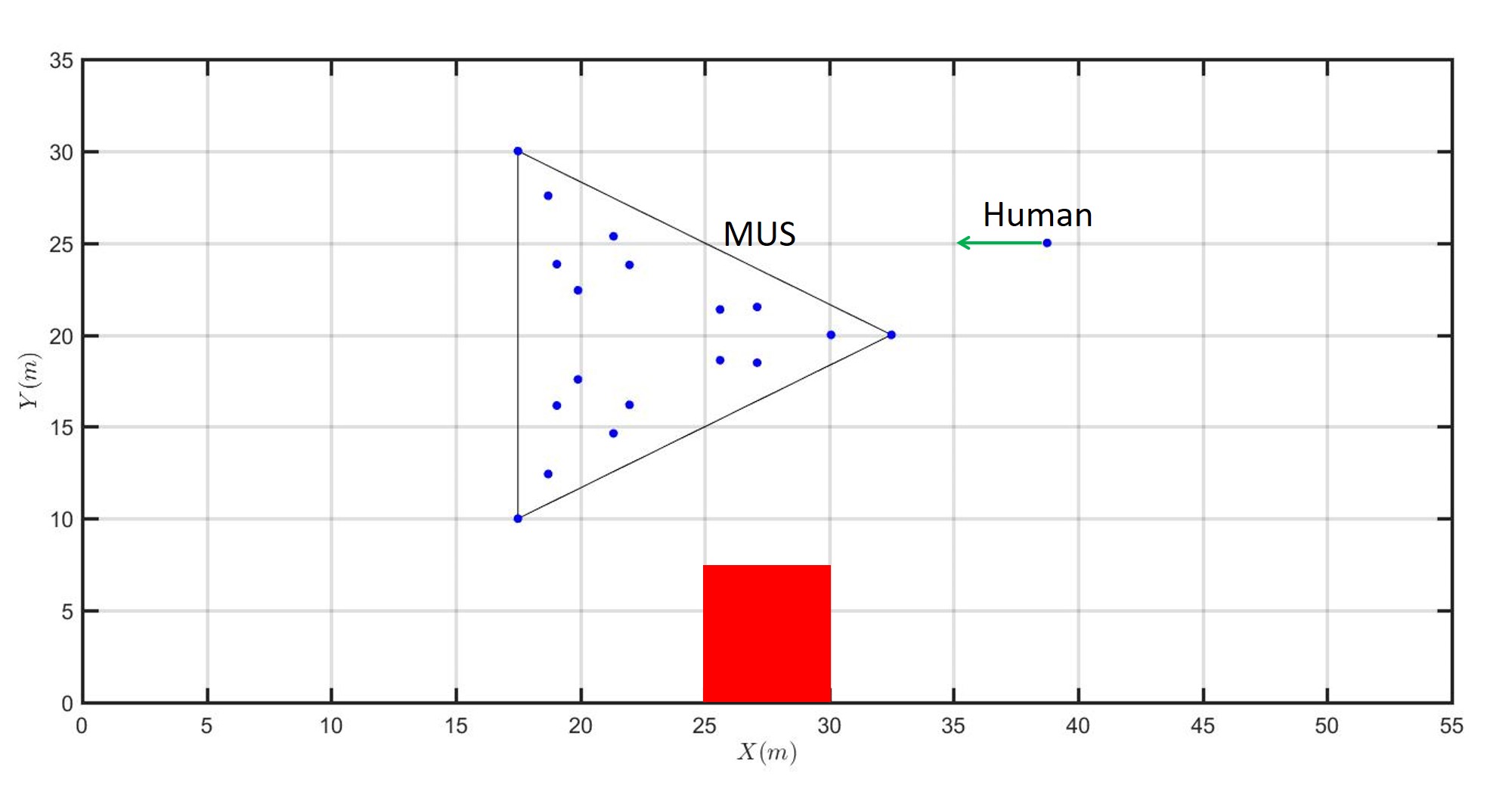}}
  \subfigure[$t=75s$]{\includegraphics[width=0.3\linewidth]{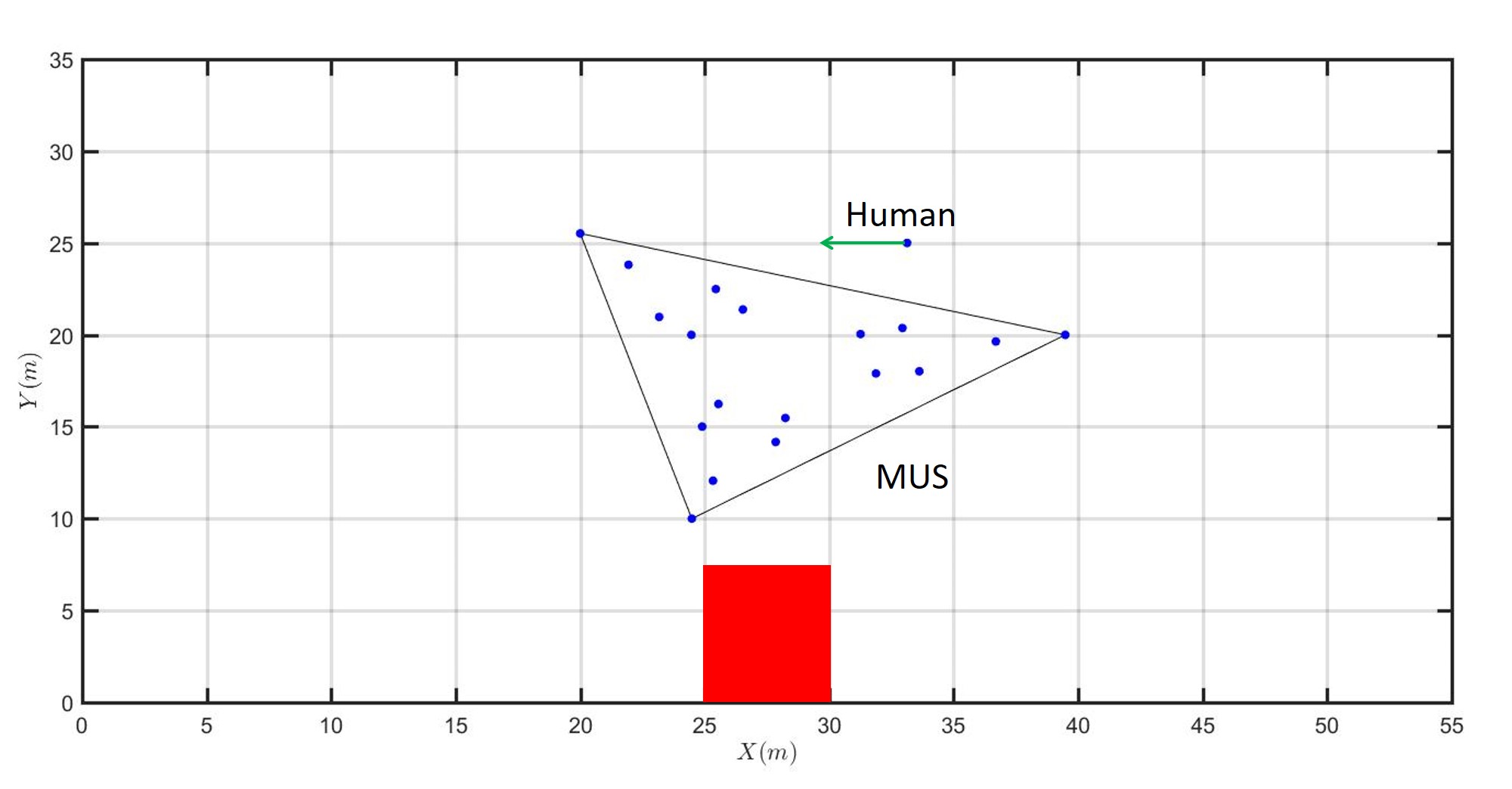}}
  \subfigure[$t=100s$]{\includegraphics[width=0.3\linewidth]{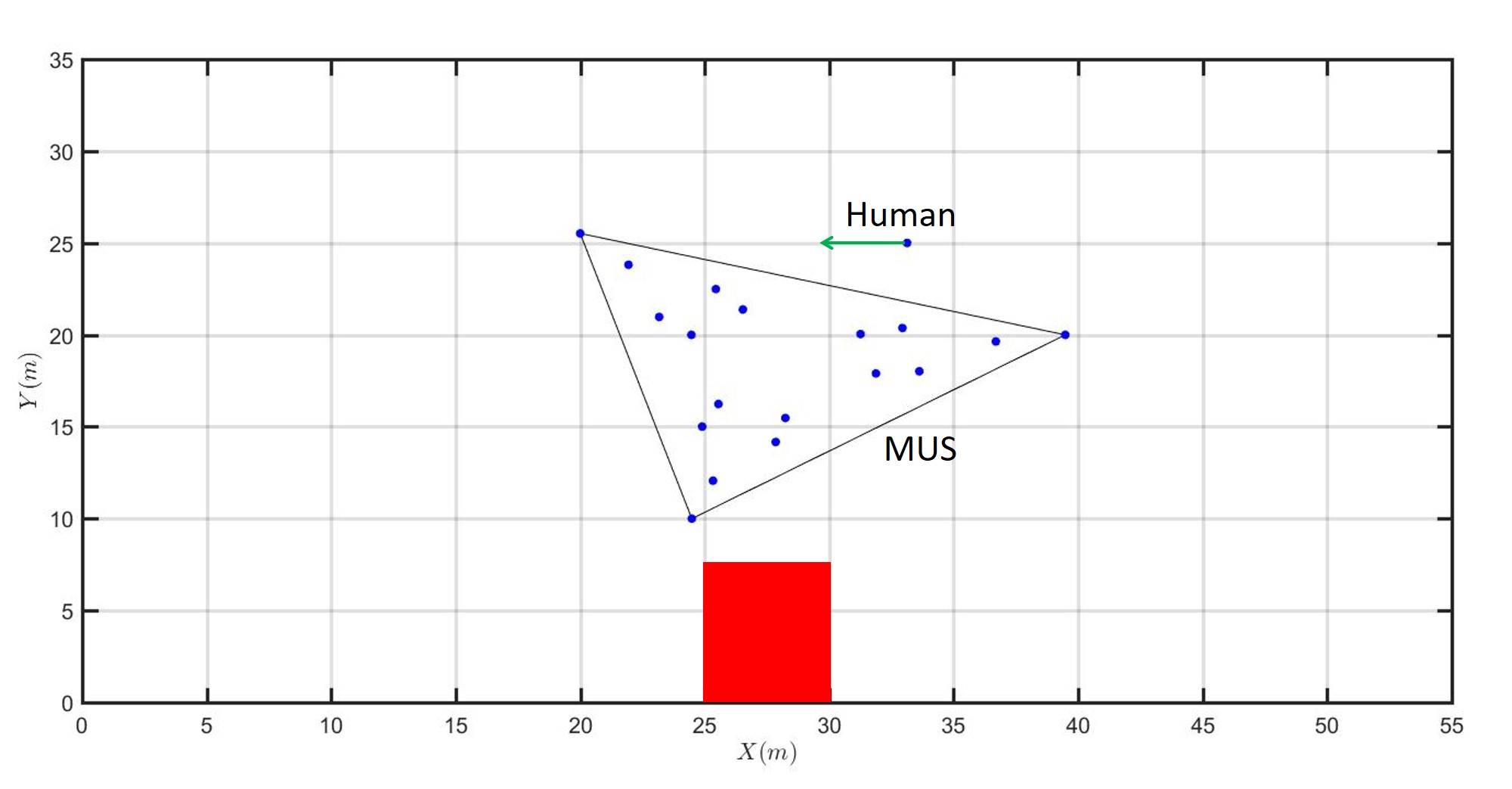}}
  \subfigure[$t=125s$]{\includegraphics[width=0.3\linewidth]{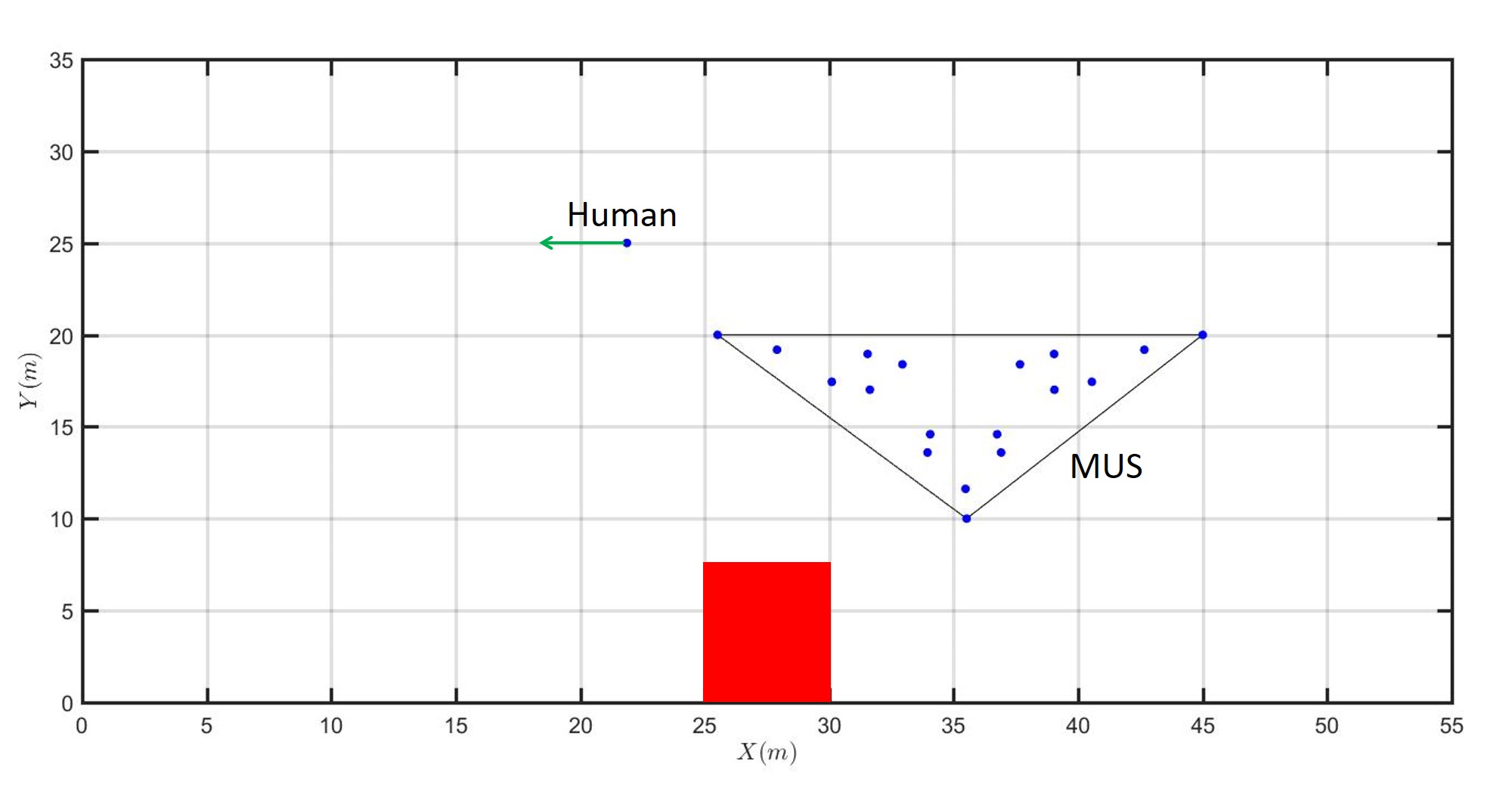}}
  \subfigure[$t=150s$]{\includegraphics[width=0.3\linewidth]{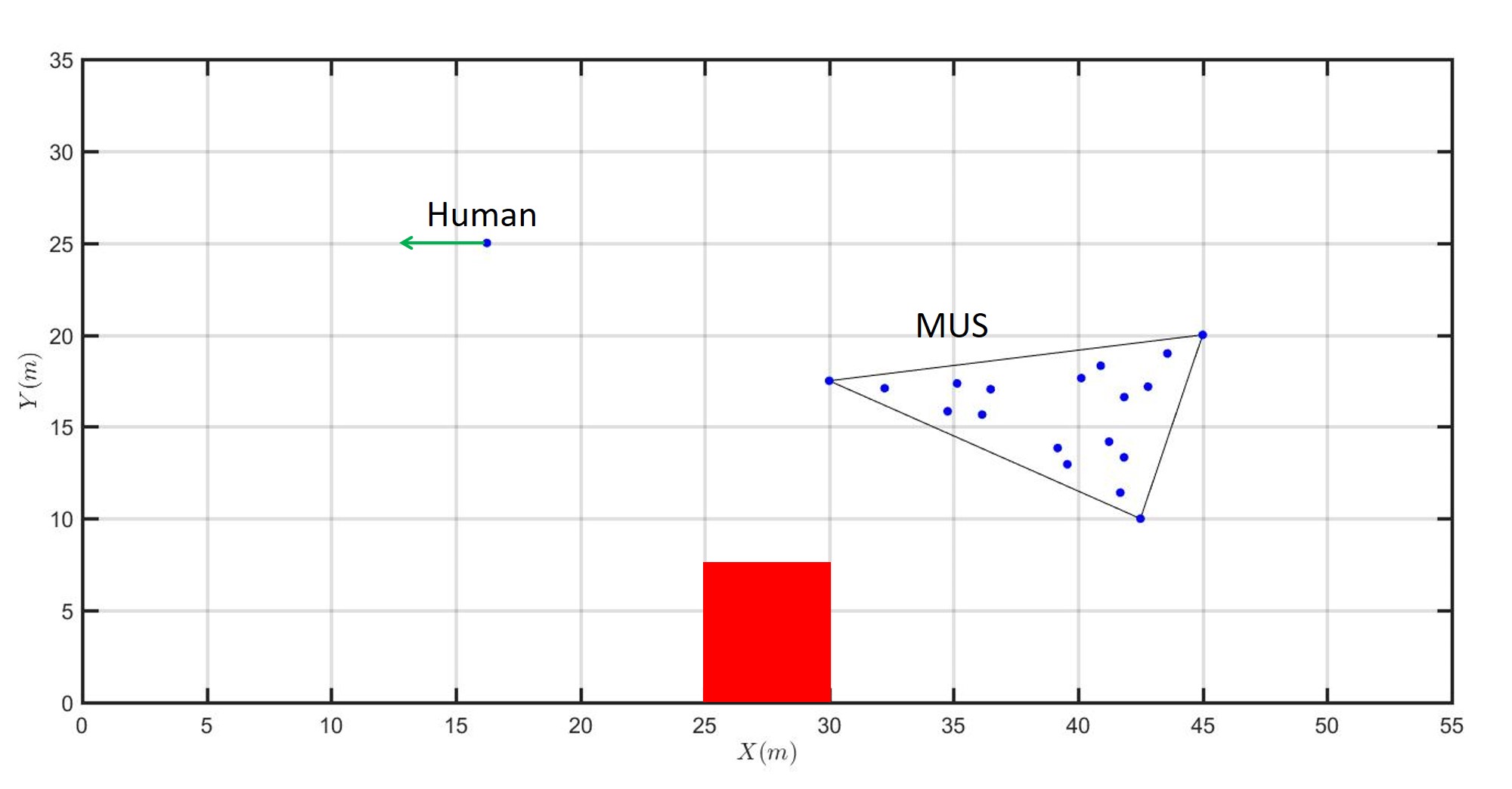}}
  \subfigure[$t=175s$]{\includegraphics[width=0.3\linewidth]{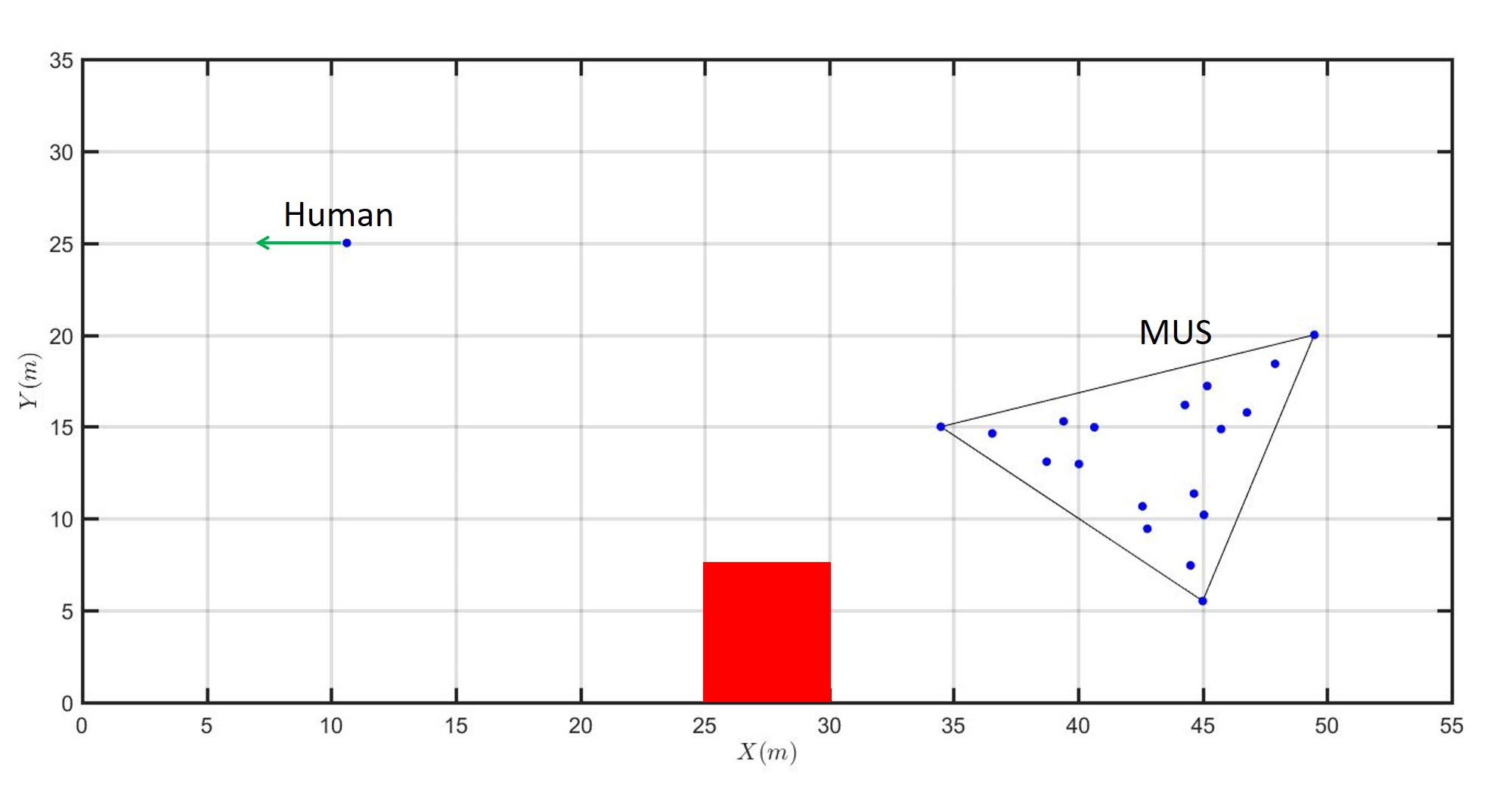}}
  \subfigure[$t=200s$]{\includegraphics[width=0.3\linewidth]{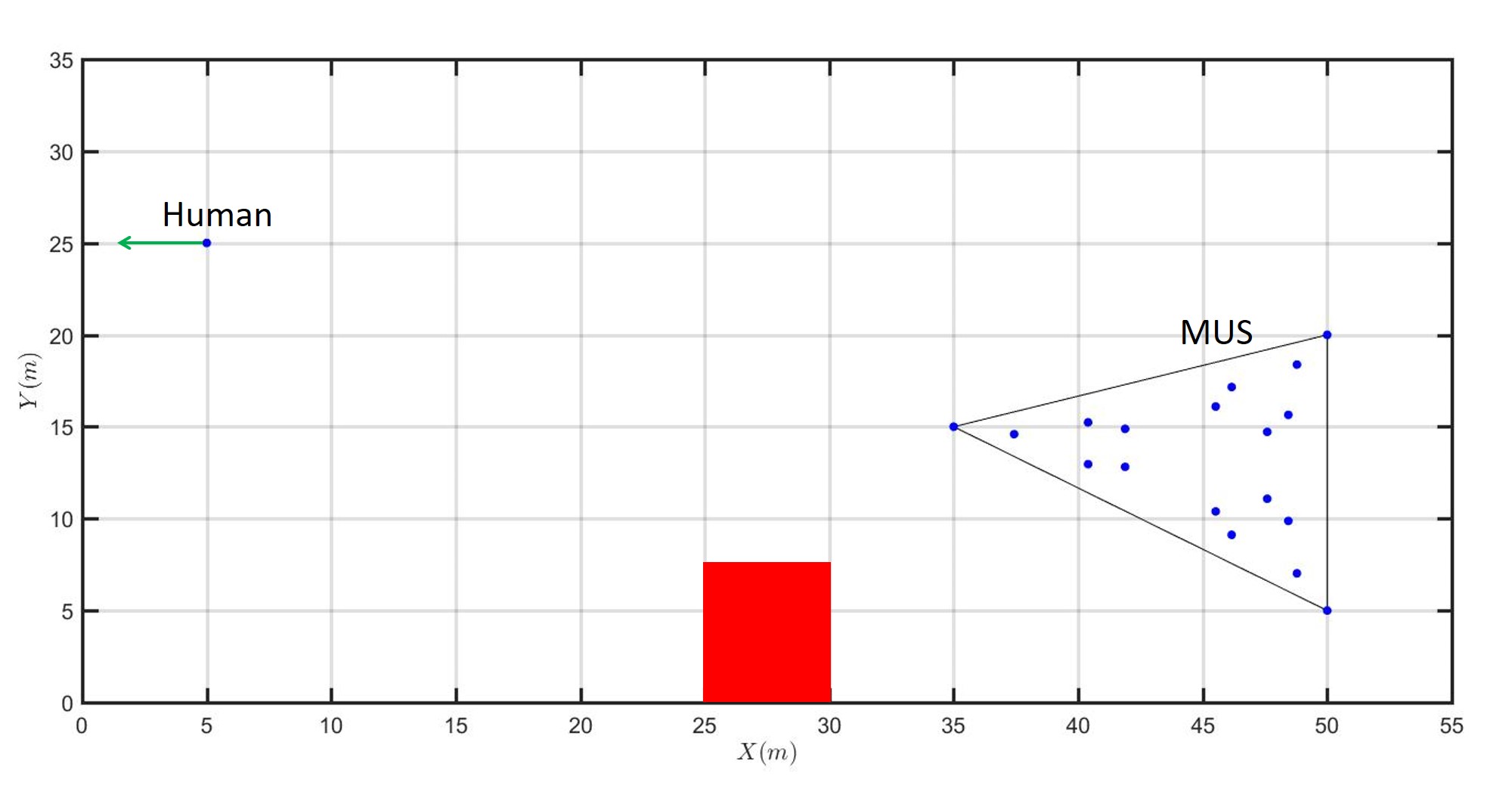}}
  \caption{(a-i) MQS-human formations in case study 2.  The isolated dot depicts the human walking across the flight area.}
\label{case22}
\end{figure}

\subsection{Case Study 2}
For the second case study, the  target MQS formation is the same as the first case study. However, initial formation is different. Leaders are initially positioned at $\mathbf{R}_{1,0}=5\hat{\mathbf{e}}_1+10\hat{\mathbf{e}}_2+10\hat{\mathbf{e}}_3$, $\mathbf{R}_{2,0}=20\hat{\mathbf{e}}_1+20\hat{\mathbf{e}}_2+10\hat{\mathbf{e}}_3$,
 $\mathbf{R}_{3,0}=5\hat{\mathbf{e}}_1+30\hat{\mathbf{e}}_2+10\hat{\mathbf{e}}_3$ (See Fig. \ref{case2initfinal}). It is assumed that a human walks from right to left with constant velocity ${45\over 200}m/s$ along the straight path 
\[
5<x<50, ~y=25.
\]
$X$ and $Y$ components of leaders' optimal trajectories and human trajectory are shown in Fig. \ref{leadeshumancase2} (a) and (b).

Shown in Fig. \ref{eigenvaluesuddeterministic} are eigenvalues of matrix $U_{\mathrm{CD}}$ associated with MQS continuum deformation in the second case study.  It is seen that $\lambda_{1}(t)=\lambda_{2}(t)=1$ ($0\leq t\leq 60s$). Therefore, the MQS moves as a rigid body over $t\in[0,60]$, while the MQS significantly deforms over $t\in (60,200]s$  Figs \ref{case22} (a)-(i) show the MQS-human configurations at different times. As shown the MQS avoids flying over the human walking from right to left.
\clearpage
\section{Conclusion}
\label{Conclusion}
This paper studies the problem of continuum deformation optimization of a UAV team flying over a populated and geometrically constrained area.   Continuum deformation is planned so that safety requirements associated with collision avoidance are satisfied and optimization cost metrics are minimized. The paper defines cost as the weighted sum of leaders' travel distances and likelihood of human presence under the flight region. 
The UAV optimization strategy proposed in this paper can be applied in a variety of missions given extensions to assure resilience given system failures. Such regions will likely contain "No-Fly-Zones", and it will be advantageous to minimize time of flight over people.

\bibliography{ifacconf}             
                                                   







\end{document}